\documentclass[draft,11pt,onefignum,onetabnum]{siamart171218}

\usepackage{lipsum}
\usepackage{amsfonts}
\usepackage{graphicx}
\usepackage{epstopdf}
\usepackage{algorithmic}
\ifpdf
  \DeclareGraphicsExtensions{.eps,.pdf,.png,.jpg}
\else
  \DeclareGraphicsExtensions{.eps}
\fi

\usepackage[utf8]{inputenc}
\usepackage{amsmath}
\usepackage{latexsym}
\usepackage{subfigure}
\usepackage{amssymb}
\usepackage{stmaryrd}
\usepackage{caption}
\usepackage{color}
\usepackage{enumerate}
\usepackage{bbm}

\newsiamremark{remark}{Remark}
\newsiamremark{hypothesis}{Hypothesis}
\crefname{hypothesis}{Hypothesis}{Hypotheses}
\newsiamthm{claim}{Claim}

\newcommand{\dfn}{\triangleq}
\newcommand{\QED}{\Box} 
\newcommand{\rw}{\rightarrow} 
\newcommand{\Prob}{\mathbb{P}}    
\newcommand{\Real}{\mathbb{R}}  
 
\newcommand{\Expect}{\mathbb{E}} 
\newcommand{\Ind}{\mathbbm{1}} 

\newcommand{\mB}{{\mathcal B}}
\newcommand{\mP}{{\mathcal P}}

\newcommand{\mD}{{\mathcal D}}

\newcommand{\mS}{{\mathcal S}}

\newcommand{\mG}{{\mathcal G}}
\newcommand{\mF}{{\mathcal F}}

\newcommand{\mX}{{\mathcal X}}
\newcommand{\mY}{{\mathcal Y}}
\newcommand{\mN}{{\mathcal N}}

\newcommand{\mfR}{\mathfrak{R}}
\newcommand{\mfM}{\mathfrak{M}}

\newcommand{\mbE}{\mathbb{E}}

\newcommand{\sff}{{\sf f}}

\newcommand{\sfc}{{\sf c}}

\newcommand{\sfk}{{\sf k}}

\newcommand{\sfr}{{\sf r}}

\newcommand{\sfs}{{\sf s}}
\newcommand{\sB}{{\sf B}}
\newcommand{\sd}{{\sf d}}

\newcommand{\cblue}{\textcolor{black}}

\newtheorem{assumption}[theorem]{\textit{Assumption}}

\headers{Stable approximation schemes...}{D. Crisan, A. L\'opez-Yela, J. Miguez}

\title{Stable approximation schemes for optimal filters\thanks{\funding{This work has been partially supported by the the Office of Naval Research (award no. N00014-19-1-2226) and the Spanish \textit{Agencia Estatal de Investigaci\'on} (awards TEC2017-86921-C2-2-R CAIMAN, TEC2015-69868-C2-1-R ADVENTURE, RTI2018-099655-B-I00 CLARA). The work of the first author has been partially supported by a UC3M-Santander Chair of Excellence grant held at the Universidad Carlos III de Madrid.}}}

\author{
Dan Crisan\thanks{Imperial College London (\email{d.crisan@imperial.ac.uk}).}
\and Alberto L\'opez-Yela\thanks{Universidad Carlos III de Madrid (\email{alyela@tsc.uc3m.es}).} 
\and Joaquin Miguez\thanks{Universidad Carlos III de Madrid (\email{joaquin.miguez@uc3m.es}).}
}

\usepackage{amsopn}

\begin{document}

\maketitle

\begin{abstract}
A stable filter has the property that it asymptotically `forgets'  initial perturbations. As a result of this property,  it is possible to construct approximations of such filters whose errors remain small in time, in other words approximations that are \emph{uniformly} convergent in the time variable.   As uniform approximations are ideal from a practical perspective, finding criteria for filter stability has been the subject of many papers. In this paper we  seek to construct  \emph{approximate} filters that stay close to a given (possibly) \emph{unstable} filter. Such filters are obtained through a general truncation scheme and, under certain constraints, are stable. The construction enables us to give a characterization of the topological properties of the set of optimal filters. In particular, we introduce a natural topology  on this set, under which the subset  of stable filters is dense.
\end{abstract}
 



\section{Introduction}

%
\subsection{State space models and optimal filters}

In this manuscript we are concerned with partially observed stochastic dynamical systems that evolve  in discrete time $t = 0, 1, 2, ...$ . Such systems consist of two stochastic processes: a {\em state} or {\em signal} process  $X=\{X_t, \ \ t\ge 0\}$ and an {\em observation} or  measurement  process $Y=\{Y_t, \ \ t> 0\}$. The states $X_t$ cannot be observed directly and the so-called {\em optimal filtering} problem \cite{Anderson79}  consists in computing, at every time instant $t$, the probability distribution of the state $X_t$ conditional on the observations $Y_1, Y_2, \ldots, Y_t$. 

The signal process $X$ is assumed to be Markovian and the observations $Y_t, \ t=1,2,... $ are assumed conditionally independent given the signal. Such systems can be fully characterised by the probability distribution of the state at time $t=0$, denoted by $\pi_0$, the Markov transition kernel that determines the probabilistic dynamics of the state  $X_t$, denoted by $\kappa_t$, and a  bounded {\em potential} function $g_t$ that relates the observation $Y_t$ with the state $X_t$. The potential function $g_t$ coincides (up to a proportionality constant) with the probability density function (pdf) of $Y_t$ conditional on $X_t$. If  $\kappa=\{ \kappa_t \}_{t\ge 1}$ is the sequence of Markov kernels and $g=\{g_t\}_{t\ge 1}$ for the sequence of bounded potentials then we can succinctly denote the system of interest as $\mS=\{\pi_0, \kappa, g\}$ and we refer to $\mS$ as a {\em state space model}. 

For a fixed sequence of observations, $Y_1=y_1, \ldots, Y_t=y_t, \ldots$, the model $\mS$ yields a deterministic sequence of probability measures $\{ \pi_t \}_{t\ge 1}$, where $\pi_t$ denotes  the probability distribution of $X_t$ conditional on the (fixed) observations $\{Y_i=y_i; i=1,\ldots, t\}$ (and the state space model $\mS$ itself).  Some features of the sequence $\{ \pi_t \}_{t\ge 1}$ are explicitly described in Section  \ref{subsectionstof} below.  If the sequence of observations $Y_1, \ldots, Y_t, \ldots$ is not fixed, but random, then the sequence $\{ \pi_t \}_{t\ge 1}$ generated by $\mS$ is random as well. In both cases,  $\{ \pi_t \}_{t\ge 1}$  is the solution to the filtering problem is the probability measure $\pi_t$, often referred to as the {\em optimal filter}  at time $t$, see e.g. \cite{Anderson79}.

Optimal filtering algorithms are procedures for the recursive computation, either exact or approximate, of the sequence $\{ \pi_t \}_{t\ge 1}$. Well known examples include the Kalman filter \cite{Kalman60} and its many variants \cite{Anderson79,Evensen03,Julier04}, or particle filters \cite{Bain08,DelMoral04,Doucet01b,Gordon93}. Such algorithms have found practical applications in a multitude of  scientific and engineering problems, including navigation and tracking \cite{Gustafsson02,Ristic04}, geophysics \cite{VanLeeuwen10}, biomedical engineering \cite{Rahni11} and many others.

%
\subsection{Stability of the optimal filter}
\label{subsectionstof}

The sequence of optimal filters $\{\pi_t\}_{t\ge 1}$ is constructed recursively by using  the Markov kernels $\kappa_t$ and the bounded potential functions $g_t$ starting with the a priori distribution $\pi_0$. For a given model $\mS=\{\pi_0, \kappa, g\}$ and a given sequence of observations, we associate a filtering operator that maps the prior $\pi_0$ to the optimal filter \cite{DelMoral04,DelMoral01c}. Let us denote this operator as $\Phi_{t|0}$, so that $\pi_t = \Phi_{t|0}(\pi_0)$ is the optimal filter at time $t$ when the initial distribution is $\pi_0$.  Assume next that we apply this operator not to the original prior distribution  $\pi_0$ but to another distribution denoted by $\tilde{\pi}_0$ and let  $\tilde \pi_t = \Phi_{t|0}(\tilde \pi_0)$ be the image of $\tilde \pi_0$.  Heuristically, we interpret $\tilde \pi_t$ as being the optimal filter with the ``wrong'' initialisation\footnote{Of course, one can also ask the question of what would happen if also the other two components $\kappa$ and $g$ that complete the triple $\mS=\{\pi_0, \kappa, g\}$ were ``wrong''. We do not discuss this question here as this is the subject of separate work.} $\tilde \pi_0 \ne \pi_0$. The optimal filter is {\em stable} when, for some properly defined metric function\footnote{Most often the total variation distance, see, e.g., \cite{DelMoral01c,Chigansky06}.} $D(\cdot,\cdot)$, we have
$$
\lim_{t\rw\infty} D( \pi_t, \tilde \pi_t ) = \lim_{t\rw\infty} D\left( \Phi_{t|0}(\pi_0), \Phi_{t|0}(\tilde \pi_0) \right) = 0.
$$ 
Let us note that stability is actually a property of the map $\Phi_{t|0}$, i.e., a property of the combination of the kernels $\kappa_t$ with the potential functions $g_t$ and the observations $y_1, \ldots, y_t$. Therefore, it would be more accurate to refer to the stability of the filtering operator $\Phi_{t|0}$ rather than the stability of the filter itself.

Stability is important both as a fundamental property of the system dynamics and for practical reasons: stable filters can, in principle, be approximated numerically with error rates that hold uniformly over time for a fixed computational effort \cite{DelMoral01c,Heine08}, while unstable filters demand that the computational complexity of the numerical approximation be increased over time in order to prevent the approximation error from growing. Heuristically, stable filters {\em forget} their initial conditions and their numerical implementations inherit this property and also progressively forget past errors, preventing their accumulation. 

The analysis of the stability of a filtering operator is not an easy task. Quoting \cite{Chigansky06} ``stability of the nonlinear filter stems from a delicate interplay of the signal ergodic properties and the observations `quality'. If one of these ingredients is removed, the other should be strengthened in order to keep the filter stable''. The authors of \cite{Chigansky06} use martingale convergence results to prove almost sure stability for sequences of integrals $\int f \sd \pi_t$, where $f$ is a test function of a particular class whose definition involves both the bounded potentials $g_t$ and the kernels $\kappa_t$ in the model \cite{Chigansky06}. Other authors resort to the analysis of the total variation distance between optimal filters obtained from different initial distributions \cite{Kleptsyna08,Douc09,VanHandel09} and relate stability to other properties of the dynamical system, often connected to the ergodicity of the state process \cite{Kleptsyna08,Douc09} or its observability and controllability (see \cite{VanHandel09} for the analysis of the continuous-time optimal filter). A recent analysis that builds upon \cite{Douc09, Kleptsyna08} but employs a different metric (which enables the inspection of integrals $\int f\sd \pi_t$ for $f$ unbounded) can be found in \cite{Gerber17}.

The main issue with the methods in \cite{Chigansky06,Kleptsyna08,Douc09,VanHandel09,Gerber17} is that stability is related to sets of conditions which are often hard to verify from the standard construction  of the filtering operator in terms of the kernels $\kappa_t$ and the potentials $g_t$. In contrast, the authors of \cite{Heine08} provide a set of relatively simple-to-verify sufficient conditions for the stability of $\Phi_{t|0}$. However, their analysis reduces to a relatively narrow class of state space models (with additive noise and exponential-family pdf's). A more general study can be found in \cite{DelMoral01c,DelMoral04}, where Dobrushin contraction coefficients \cite{Dobrushin56,Dobrushin70} are used as the key tool to obtain conditions on $\kappa_t$ and $g_t$ which are sufficient for stability.

To the best of our knowledge, there has been no attempt to  obtain a topological characterisation of stable filters. Rather natural questions, such as whether stable filters are ``many'' or ``few'' for a given class of state space models have not been investigated to-date.

%
\subsection{Contributions}

Let $\mS$ be a state space model with an associated sequence of filtering operators $\Phi_{t|0}$. For a given prior $\pi_0$, the latter operators yield the sequence of optimal filters $\pi_t=\Phi_{t|0}(\pi_0)$. In this paper we investigate truncation methods to construct approximate state space models and operators, denoted $\mS^n$ and $\Phi_{t|0}^n$, respectively, such that the resulting approximate filters $\pi_t^n=\Phi_{t|0}^n(\pi_0)$ 
\begin{itemize}
\item[(a)] can be guaranteed to be stable and 
\item[(b)] converge to the optimal filters, i.e., $\pi_t^n \rw \pi_t$ as $n\rw\infty$.
\end{itemize}

We tackle this two-fold problem in several steps. First, we impose a topology $\mD$ on the space of state space Markov models. Convergence of a sequence of models in this topology, i.e., $\lim_{n\rw\infty} \mS^n = \mS$, implies convergence of the associated filters, i.e., $\lim_{n\rw\infty}\pi_t^n=\pi_t$. While other approaches may be feasible, we focus on sequences of {\em truncated} models. For a given integer $n$, choose a sequence of subsets of the state space $\{C_t^n\}_{t\ge 1}$. The truncated version of a model $\mS=\{\pi_0,\kappa,g\}$ is obtained by truncating the (bounded) potentials (i.e., by constructing new functions $g_t^n$ such that $g_t^n(x)=g_t(x)$ if $x \in C_t^n$ and $g_t^n(x)=0$ otherwise). The truncated model denoted by $\mS^n=\{\pi_0,\kappa,g^n\}$, where $g^n= \{g_t^n\}_{t\ge 1}$ but the Markov kernels $\kappa$ are left unaltered. For this construction it is straightforward to verify that $\mS^n$ converges to $\mS$ in a topology $\mD$ described below and, therefore, $\pi_t^n \rw \pi_t$ when $n\rw\infty$ as well. 

One of the main contributions of this paper is to identify a broad class of state space models, denoted $\mfR$, and a procedure for the construction of the sets $\{C_t^n\}_{t\ge 1}$, which guarantees that, for every $\mS \in \mfR$ and every integer $n$, the truncated model $\mS^n$ yields a stable sequence of filters $\pi_t^n = \Phi_{t|0}^n(\pi_0)$. We also show, by way of an example, that the class of models $\mfR$ contains unstable filters and illustrate in detail how the proposed technique can be put to work in order to obtain stable approximations of these unstable filters.

In the last part of the paper we investigate more elaborate approximation schemes that involve not only the truncation of the potentials $g_t$ but also the {\em modification of the Markov kernels} $\kappa_t$ according to the sequence of sets $\{C_t^n\}_{t\ge 1}$. Let the new approximation of $\mS$ be denoted $\tilde \mS^n$. We prove that
\begin{itemize}
\item[(i)] $\tilde \mS^n$ converges to $\mS$ in the topology $\mD$ and, therefore, the resulting filters $\tilde \pi_t^n$ converge to $\pi_t$ as well;
\item[(ii)] if the sets $\{C_t^n\}$ contain a sufficiently large probability mass, then  $\lim_{n\rightarrow \infty}\sup_{t}D_{tv}(\tilde \pi_t^n , \pi_t) =0$, in other words $\pi_t^n$ converges to $\pi_t$ {\em uniformly over time}. Here $D_{tv}$ denotes the total variation distance. 
\end{itemize}
Finally, we provide explicit conditions on the potentials $g$ and the Markov kernels $\kappa$ to guarantee that the posterior probability measure of the sets $\{C_t^n\}_{t\ge 1}$ can be made large enough to ensure the uniform convergence over time of the approximate filters.

%
\subsection{Organisation of the paper}

We complete the introduction with a brief summary of the notation used through the manuscript, incorporated in Section \ref{ssNotation}. Section \ref{sBackground} covers the description of the optimal filtering problem for state space Markov models and a formal definition of the notion of stability for sequences of optimal filters. In Section \ref{sTruncated} we introduce the proposed truncation method and in Section \ref{sASSMs} we provide regularity constraints sufficient to guarantee the stability of the resulting approximate filters. Section \ref{sExample} contains an example of an unstable filter that can be obtained as the limit of a sequence of stable filters. In Section \ref{sUniform} we introduced the enhanced truncation scheme and show that it can yield approximate filters that converge uniformly over time. Finally, in Section \ref{sConclusions} we make some brief concluding remarks.

%
\subsection{Notation} \label{ssNotation}

We summarise the notation used throughout the manuscript, roughly organised by topics.

\begin{itemize}
\item Sets, measures and integrals: 
        \begin{itemize}
        \item $\mB(S)$ is the $\sigma$-algebra of Borel subsets of $S \subseteq \Real^d$.
        \item $\mP(S) := \{ \mu: \mB(S) \mapsto [0,1] \mbox{ and $\mu(S)=1$} \}$ is the set of probability measures over $\mB(S)$.
        \item $(f,\mu) \dfn \int f \sd \mu$ is the integral of a Borel measurable function $f:S \mapsto \Real$ with respect to the measure $\mu \in \mP(S)$.
        \item  The indicator function on a set $S$ is denoted $\Ind_S(x)$.
        Given a measure $\mu$ and a set $S$ we equivalently denote $\mu(S):=(\Ind_S,\mu)$.
        \item Let $A$ be a subset of a reference space $\mX \subset \Real^d$. The complement of $A$ with respect to $\mX$ is denoted $\bar A := \mX \backslash A$. 
        \end{itemize}

\item Functions and sequences:
        \begin{itemize}
        \item $\sB(S)$ is the set of bounded Borel measurable real functions over $S$. Given a sequence $\{ f_t \in \sB(S) \}_{t\ge 1}$, we denote 
        $$
        \| f_t \|_\infty := \sup_{s \in S} |f(s)| \quad \mbox{and} \quad  
        \| f \|_\infty := \sup_{t\ge 1} \| f_t \|_\infty.
        $$
        \item We use a subscript notation for finite subsequences, namely $x_{t_1:t_2} \dfn \{ x_{t_1}, \ldots, x_{t_2} \}$.
        \end{itemize}

\item Random variables:
        \begin{itemize}
        \item $\Expect[\cdot]$ denotes expectation with respect to the underlying probability measure $\Prob$ when working on the probability space  $\{ \Omega, \mathcal F, \Prob\}$. 
        \item Random variables are denoted by capital letters (e.g., $Z:\Omega \mapsto \Real^d$) and their realisations by lower case letters (e.g., $Z(\omega)=z$ or, simply, $Z=z$).
        \item If $Z$ is a r.v. taking values in $S \subseteq \Real^d$, with probability distribution  $\mu \in \mP(S)$, then the $L_p$ norm of $Z$ is given by $\| Z \|_p := \Expect\left[|Z|^p\right]^{\frac{1}{p}} = \left( \int |z|^p \mu(dz) \right)^{\frac{1}{p}}$, $p \ge 1$. 
        \end{itemize}

\end{itemize}

\section{State space models and optimal filters} \label{sBackground}

\subsection{Markov state-space models in discrete time}

Let $\{ \Omega, \mathcal F, \Prob\}$ be a probability space, where $\Omega$ is the sample space, $\mF$ is a  $\sigma$-algebra and $\Prob$ is a probability measure on which we consider two stochastic processes:
\begin{itemize}
\item the {\em signal} or {\em state} process $X=\{ X_t \}_{t \ge 0}$, with values in the space $\mX \subseteq \Real^{d_x}$, 
\item the {\em observation} process $Y=\{ Y_t \}_{t \ge 1}$, with values in the space $\mY \subseteq \Real^{d_y}$. 
\end{itemize}

We assume that the state process evolves over time according to the family of Markov kernels 
\begin{equation}
\kappa_t(A | x_{t-1}) = \Prob\left( X_t \in A | X_{t-1} = x_{t-1} \right),
\nonumber
\end{equation} 
where $A \in \mB(\mX)$ and $x_{t-1}\in\mX$. The observation process is described by the conditional distribution of the observation $Y_t$ given the state $X_{t}$. Specifically, we assume that  $Y_t$ has a conditional pdf $g_t(y_t|x_t)$ w.r.t. a reference measure $\lambda$ (usually, but not necessarily, the Lebesgue measure), given the state $X_t=x_t$. The observations are assumed to be conditionally independent given the states.

If the sequence $\{ Y_t = y_t \}_{t \ge 1}$ is fixed, then we write $g_t(x_t) := g_t(y_t|x_t)$ for conciseness and to emphasise that $g_t$ is a function of the state $x_t$, i.e., we use $g_t(x)$ as the potential of $x \in \mX$ given the observation $y_t$. When the observation sequence is random, we write $g_t^{Y_t}(x)=g_t(Y_t|x)$ for the potential of $x\in\mX$. Note that, for fixed $x\in\mX$, $g_t^{Y_t}(x)$ is a r.v. itself.


\cblue{The state process $X_t$ (with prior probability law $\pi_0(\sd x_0)$ and Markov transition kernels $\kappa_t(\sd x_t|  x_{t-1})$) and the observation process $Y_t$ (related to $X_t$ by means of the pdf's $g_t(y_t|x_t)$) yield the typical formulation of a state space Markov model. In this paper, we use the term state space model to refer to the triple $\mS=\{\pi_0, \kappa, g\}$, where $\kappa=\{\kappa_t\}_{t\ge 1}$ is the family of Markov kernels for the process $X_t$ and  $g=\{g_t(y_t|\cdot)\}_{t\ge 1}$ is the family of potentials generated by the observations $\{Y_t=y_t\}_{t \ge 1}$. This is a slight abuse of the usual terminology. However, as will be shown in Section \ref{subsection_of} below, the triple $\mS$ contains all the necessary ingredients needed to specify the conditional probability law of the state $X_t$ given the observations $Y_{1:t}=y_{1:t}$, for every $t \ge 1$. These conditional probability distributions are the main object of this paper, and hence we assimilate $\mS$ to the state space model itself.}

\subsection{The optimal filter}
\label{subsection_of}

The filtering problem consists in the computation of the posterior probability measure of the state $X_t$ given a sequence of observations up to time $t$. Specifically, we aim at the sequence of probability measures
\begin{equation}
\pi_t(A) := \Prob\left( X_t \in A | Y_{1:t}=y_{1:t} \right), \quad t \ge 1,
\nonumber
\end{equation}
where $A \in \mB(\mX)$. The measure $\pi_t$ is commonly called the {\em optimal filter} at time $t$. 
 $\pi_t$ is computed from $\pi_{t-1}$ in two steps. First, we obtain the {\em predictive} probability measure 
\begin{equation}
\xi_t(A) := \Prob\left( X_t \in A | Y_{1:t-1}=y_{1:t-1} \right)
\nonumber
\end{equation}
and then we compute $\pi_t$ from $\xi_t$. To be precise, we write $\xi_t = \kappa_t\pi_{t-1}$, meaning that\footnote{Here and in all subsequent formulae $f:\mathcal X\rw\mathbb{R}$ is a bounded measurable function.}
$$
(f,\xi_t) = (f,\kappa_t\pi_{t-1}) = ((f,\kappa_t),\pi_{t-1}),
$$
 and $\pi_t = g_t \cdot \xi_t$, which is defined as
\begin{equation}
(f,g_t\cdot \xi_t) := \frac{
        (fg_t,\xi_t)
}{
        (g_t,\xi_t)
}.
\label{eqProjective}
\end{equation}

The definitions above are given for a fixed (but arbitrary, unless otherwise stated) sequence of observations $Y_{1:\infty}=y_{1:\infty}$. In this case, the state space model $\mS=\{\pi_0, \kappa, g\}$  yields deterministic sequences $\{\pi_t\}_{t\ge 1}$ and $\{ \xi_t \}_{t\ge 1}$. If the observations are random, then the model $\mS=\{\pi_0,\kappa, g^Y\}$, where $g^Y:=\{g_t^{Y_t}\}_{t\ge 1}$, yields sequences of {\em random} measures $\pi_t^{Y_{1:t}}$ and $\xi_t^{Y_{1:t-1}}$, $t \ge 1$.

%
\subsection{The prediction-update operator}\label{PU_sect}

The transformation of the filter $\pi_{t-1}$ into $\pi_t$ can be represented by the composition of two operators $\Psi_t, \Upsilon_t : \mP(\mX)\mapsto\mP(\mX)$ which can be defined as follows:
\begin{itemize}
\item the prediction (P) operator $\Psi_t(\mu) := \kappa_t\mu$, where $\mu\in\mP(\mX)$, yields
$$
(f,\Psi_t(\mu) ) := (f,\kappa_t\mu) = ((f,\kappa_t),\mu),
$$

\item and the update (U) operator yields $\Upsilon_t(\mu) := g_t \cdot \mu$, i.e., 
$$
(f,\Upsilon_t(\mu)) := \frac{
        (fg_t,\mu)
}{
        (g_t,\mu)
}.
$$
\end{itemize}
By composing the maps $\Psi_t$ and $\Upsilon_t$ we obtain the prediction-update (PU) operator
$\Phi_t (\mu) := (\Upsilon_t \circ \Psi_t)(\mu)$
such that
\begin{equation}
(f,\Phi_t(\mu))=  (f,\Upsilon_t\left( \Psi_t(\mu) \right)
= \frac{
        (fg_t,\kappa_t\mu)
}{
        (g_t,\kappa_t\mu)
} = (f, g_t \cdot \kappa_t \mu),
\label{eqOperators2}
\end{equation}
which obviously implies $\pi_t = \Phi_t(\pi_{t-1})$. If we additionally denote the composition of PU operators as
$$
\Phi_{t|k} := \Phi_t \circ \Phi_{t-1} \circ \cdots \circ \Phi_{k+1}
$$
then we can compactly represent the evolution of the filter over $t-k$ consecutive steps, namely $\pi_t = \Phi_{t|k}(\pi_{k})$. Note that the map $\Phi_t$ depends on the Markov kernel $\kappa_t$ and the likelihood $g_t$ alone (and not on the prior measure $\pi_0$). 


\subsection{A topological structure on the set of state space Markov models}

Let $\mfM$ be the set of state space Markov models of the form $\mS=\{ \pi_0, \kappa, g \}$ \cblue{for which the sequence of filters $\{ \pi_t \}_{t\ge1}$ is well defined. In particular, we assume that the normalisation constant $(g_t,\xi_t)$ in Eq. \eqref{eqProjective} is strictly positive\footnote{
\cblue{This is ensured, for example, if the two conditions below are satisfied: 
\begin{itemize}
\item[(i)] There exists $S \subseteq \mX$ such that $g_t(x)>0$ (strictly positive) for every $x\in S$ and $\text{Leb}(S)>0$, i.e., the subset $S$ has positive Lebesgue measure.
\item[(ii)] The transition kernel $\kappa_t$ puts positive probability mass on all subsets with positive Lebesgue measure, i.e., for every $x \in \mX$ and every $S' \subseteq \mX$ such that $\text{Leb}(S')>0$, we have $\kappa_t(S'|x)>0$.
\end{itemize}
}
} for all $t \ge 1$, in order to guarantee that $\pi_t \in \mP(\mX)$ and, therefore, $| (f,\pi_t) | < \infty$ whenever $f \in B(\mX)$.}

It would be tempting to impose a topological structure based on the convergence of the fixed-dimensional marginal distributions  $\pi_t= \Phi_{t:0}(\pi_0)$, $t \ge 1$. This structure, however, would be misleading because there is no one-to-one correspondence between elements of $\mfM$ and their corresponding probability distributions $\{\Phi_{t:0}(\pi_0)\}_{t\ge 1}$.  In particular, for every state space model $\mS\in\mfM$ one can construct a related model $\mathcal{S}'\in\mfM$, with PU operators $\Phi_t^\prime$, such that $\Phi_{t|0}^\prime(\pi_0)=\Phi_{t|0}(\pi_0)$ and $\Phi_{t|0}^\prime$ is stable, while $\Phi_{t|0}$ is not\footnote{For example, consider the models $\mS=\{\pi_0,\kappa,g\}$ and $\mS' = \{\pi_0,\kappa',g\}$, where $\kappa_t'(\sd x_t|x_{t-1}) = \int_\mX \kappa_t(\sd x_t|\bar x_{t-1}) \pi_{t-1}(\sd \bar x_{t-1}) = \xi_t(\sd x_t)$. It is apparent that $\pi_t' = g_t \cdot \kappa_t' \pi_{t-1}' = g_t \cdot \xi_t = \pi_t$.}. 
For this reason, we impose a topology $\mD$ directly on the components of the state space models. To be specific: 

\begin{itemize}

\item We endow $\mP(\mathcal{X})$ with the metrisable topology given by the total variation distance 
\[
D_{tv}(\alpha,\beta) := \sup_{A} | \alpha(A)-\beta(A) |,\ \  \alpha,\beta\in\mP(\mathcal{X}),
\]
 where the supremum is taken over all measurable sets.

\item The sequence of Markov kernels $\kappa^{n}=\{ \kappa_t^n \}_{t \ge 1}$ converges to $\kappa = \{ \kappa_t \}_{t\ge 1}$ when  
\begin{equation}\label{deftopk}
\lim_{n\rw\infty} D_{tv}(\kappa_t^n(\cdot,x),\kappa_t(\cdot,x))=0, \quad  \mbox{for every $t\ge 1$ and any $x\in\mathcal X$,}
\end{equation}
and we denote $\lim_{n\rw\infty} \kappa^n = \kappa$.

\item  We impose the topology of bounded convergence on the set of (\cblue{non-negative}) bounded potential functions. More precisely, we say that the sequence $g^{n}=\{g_t^n\}_{t\ge 1}$ is uniformly bounded when there exists $G<\infty$ such that $\sup_{n,t\ge 1} \|g_t^n\|_\infty < G$. Then, a uniformly bounded sequence $g^n$ converges to $g=\{g_t\}_{t\ge 1}$ when  
\begin{equation}\label{deftopg}
\lim_{n\rw\infty}g_{t}^n(x)=g_t(x), \quad \mbox{for every $t\ge 0$ and any $x\in\mathcal X$,}
\end{equation} 
and we write $\lim_{n\rw\infty} g^n = g$.

\item Finally, a sequence of state space models $\mS^n=\{ \pi_0^n, \kappa^n, g^n \}$ converges to the model $\mS=\{ \pi_0, \kappa, g \}$ in the topology $\mD$ when $\lim_{n\rw\infty} D_{tv}(\pi_0^n,\pi_0)=0$, $\lim_{n\rw\infty} \kappa^n = \kappa$ and $\lim_{n\rw\infty} g^n = g$. We denote $\lim_{n\rw\infty} \mS^n = \mS$.
\end{itemize} 

\begin{remark}  A complete description of the topology $\mD$ described above can be outlined as follows. A generator of open sets for the first component of the topology is given by the open balls
 \begin{equation}
 B(\pi_0, r) = \left\{ 
        \pi \in \mP(\mathcal{X}) : D_{tv}(\pi,\pi_0)<r
\right\}, \quad \mbox{for $\pi_0\in P (\mathcal  X)$ and $r>0$}.
\nonumber
 \end{equation}
 The topology on the second component is defined to be the smallest topology such that all functions $f_{x,t}:\mfM \mapsto \mathcal P (\mathcal  X)$, $x\in \mathcal X$, defined as $f_{x,t}(\{\pi_0, \kappa, g\})=\kappa_t(\cdot|x)$ are continuous. The topology on the third component is the smallest topology such that all functions $h_{x,t}:\mfM\mapsto \mathbb R$, $x\in \mathcal X$, defined as $h_{x,t}(\{\pi_0, \kappa, g\})=g_t(x)$ are continuous when restricted to  $\mfM_G:=\left\{ \{ \pi_0, \kappa, g \} \in \mfM : \sup_{t\ge 1} \|g_t\|_\infty<G \right\}$ for any $0<G<\infty$. This topology  is not metrisable: \cblue{while convergence for sequences $\pi_0^n \in \mP(\mX)$ can be expressed in terms of the total variation distance, neither the convergence of $\kappa^n$ nor the convergence of $g^n$ can be recast in terms of proper metrics. Recall that $\kappa$ is a family of kernels and $g$ is a family of potentials, both indexed by $(x,t) \in \mX \times \mathbb{Z}^+$.}
 \end{remark}

\begin{remark} If we restrict $\mfM$ to the set of state space models for which the initial probability measure and corresponding kernels are absolutely continuous with respect to a fixed reference measure $\lambda$, then then we can relax the limits in \eqref{deftopk} and \eqref{deftopg} to hold $\lambda$-almost surely.
\end{remark}

\begin{remark} \label{rmMetric} By imposing uniform convergence (over the time and the state variables) on the set of kernels and the space of potential functions we can introduce a slightly stronger topology $\mD'$ on $\mfM$ that has the advantage of being metrisable. To be specific, we define the distance between two state space models $\mS=\{\pi_0, \kappa, g\}$ and $\mS'=\{\pi_0', \kappa', g'\}$ as
\begin{equation}
D_{\mfM}(\mS,\mS') := D_{tv}(\pi_0,\pi_0')+\sup_{t\ge 1}\sup_{x\in\mX}D_{tv}(\kappa_t(\cdot|x),\kappa_t'(\cdot|x))
+\sup_{t\ge 1}\sup_{x\in\mathcal X}|g_t(x)-g_t'(x)|.
\label{defMetric_in_S}
\end{equation}   
\end{remark}

The topology $\mD$ has the property that convergence of the sequence of models $\mS^n$, $n\ge 0$, to $\mS$ implies convergence of the marginal probability measures $\pi_t^n$ (generated by the models $\mS^n$) towards the optimal filter $\pi_t$ generated by model $\mS$. This result is made rigorous by the following lemma.

\begin{lemma} \label{lmMarginals} Let $\mS^n=\{\pi_0^n,\kappa^n,g^n\}$, $n\ge 0$, and $\mS=\{\pi_0,\kappa,g\}$ be elements of $\mfM$ with corresponding PU operators $\Phi_t^n$  and $\Phi_t$, respectively. If $\lim_{n \rw \infty} \mS^n=\mathcal S$, then $\lim_{n \rw \infty} \Phi_{t|0}^n(\pi_0^n) = \Phi_{t|0}(\pi_0)$.    
\end{lemma}    
\begin{proof}
We proceed with a standard induction argument. The case $t=0$ holds trivially, since $\lim_{n\rw\infty} \mS^n=\mS$ implies that $\lim_{n\rw\infty} D_{tv}(\pi_0^n,\pi_0)=0$. For the induction step, assume that $\lim_{n \rw \infty} D_{tv}\left( \beta_n, \beta \right) = 0$ for any $t\ge1$, where $\beta_n=\Phi_{t-1|0}^n(\pi_0^n)$ and $\beta=\Phi_{t-1|0}(\pi_0)$. As we apply the prediction operator $\Psi_t^n(\alpha)=\kappa_t^n\alpha$, a straightforward triangle inequality yields
\begin{eqnarray}
|(f,\Psi_t^n( \beta^n))-(f,\Psi_t\left( \beta\right))| &=& \nonumber 
\left|
        \left(
                f,\Psi_t^n\left(\beta^n\right)
        \right) - \left(
                f,\Psi^{n}_t\left( \beta\right)
        \right)+
        \left(
                f,\Psi_t^n\left( \beta\right)
        \right) - \left(
                f,\Psi_t\left( \beta\right)
        \right)
\right|\\
&\le&   
\left|
        \left( 
                (f,\kappa^n), \beta^n-\beta
        \right)
\right| + 
\left|
        \left(
                (f,(\kappa^n-\kappa)), \beta
        \right)
\right| \nonumber \\
&\le& \nonumber 
\| f \|_{\infty} D_{tv}\left(
        \beta^n, \beta
\right) +
\left|
        \left(
                (f,(\kappa^n-\kappa)), \beta
        \right)
\right|,\nonumber \\&&
\label{eqPstep}
\end{eqnarray}
where the last inequality follows from the definition of total variation distance. The first term on the right hand side of \eqref{eqPstep} converges to zero by the induction hypothesis, while the second term converges to zero by the bounded convergence theorem \cite{Williams91}.

Next, we write the PU operator $\Phi_t^n$ in terms of the P operator $\Psi_t^n$ to obtain 
\begin{eqnarray}
\left|
        \left(
                f,\Phi_t^n\left(\beta^n \right)
        \right) - \left(
                f,\Phi_t\left( \beta\right)
        \right) 
\right| &=& \nonumber 
\left|
        \frac{
                \left(
                        fg^{n}_t, \Psi_t^n\left(\beta^n \right)
                \right)
        }{
                \left(
                        g^{n}_t, \Psi_t^n \left(\beta^n\right)
                \right)
        } - \frac{
                \left(
                        fg_t,\Psi_t\left( \beta\right)
                \right)
        }{
                \left(
                        g_t,\Psi_t \left( \beta\right)
                \right)
        }
\right|\\
 &\le& \nonumber 
\left|
        \frac{
                \left(
                        fg^{n}_t, \Psi_t^n \left( \beta^n \right)
                \right)
        }{
                \left(
                        g^{n}_t,\Psi_t^n \left( \beta^n\right)
                \right)
        } - \frac{
                \left(
                        fg^n_t, \Psi_t^n \left( \beta^n)\right)
                \right)
        }{
                \left(
                        g_t, \Psi_t \left( \beta\right)
                \right)
        }
\right| 
\\&&+ \nonumber  
\left|
        \frac{
                \left(
                        fg^{n}_t, \Psi_t^n \left( \beta^n \right)
                \right)
        }{
                \left(
                        g_t, \Psi_t \left( \beta \right)
                \right)
        } - \frac{
                \left(
                        fg_t,\Psi_t\left( \beta \right)
                \right)
        }{
                \left(
                        g_t, \Psi_t \left( \beta\right)
                \right)
        }
\right| \\
&\le& \nonumber  
\frac{
        \|f\|_{\infty} \left|
                \left(
                        g^{n}_t, \Psi_t^n \left( \beta^n\right)
                \right) - \left(
                        g_t, \Psi_t \left( \beta\right)
                \right)
        \right|
}{
        \left(
                g_t,\Psi_t \left( \beta\right)
        \right) 
} \\ 
 && +\frac{
        \left|
                \left(
                        fg^{n}_t, \Psi_t^n \left( \beta^n\right)
                \right) - \left(
                        fg_t, \Psi_t \left( \beta \right)
                \right)
        \right|
}{
        \left(
                g_t, \Psi_t\left( \beta \right)
        \right) 
}.
\label{eqLarge}
\end{eqnarray}
$\left.\right.$\\[-5mm]
However, inequality \eqref{eqPstep} implies that both terms on the right hand side of \eqref{eqLarge} converge to 0, hence the proof is complete.
\end{proof}

\begin{remark} Although the topology $\mD'$ is stronger than the topology $\mD$, it does not imply the uniform convergence (over time) of the corresponding PU-operators.  In particular, Lemma \ref{lmMarginals} ensures that for any $\epsilon>0$ and any $t<\infty$ there exists $n_{\epsilon,t}\ge 0$ such that $D_{tv}( \Phi_{t|0}^n(\pi_0), \Phi_{t|0}(\pi_0) ) < \epsilon$ whenever $n \ge  n_{\epsilon,t}$. However, the lemma does {\em not} guarantee that $\sup_{t\ge 1}  D_{tv}( \Phi_{t|0}^n(\pi_0), \Phi_{t|0}(\pi_0) ) < \epsilon$ for any finite value of $n$.
\end{remark}


\subsection{Stability of the optimal filter} \label{ssStab}

The sequence of optimal filters $\{\pi_t\}_{t\ge 1}$ generated by a state space Markov model $\mS=\{\pi_0,\kappa,g\}$ is stable when the dependence of $\pi_t$ on the prior measure $\pi_0$ vanishes over time. Formal definitions are provided next.

\begin{definition} \label{defStability}
Let $\{ \Phi_t \}_{t\ge 1}$ be a sequence of PU operators defined on $\mP(\mX)$. The sequence of optimal filters $\{ \pi_t \}_{t\ge 1}$ generated by $\{ \Phi_t \}_{t\ge 1}$ is stable when
$$
\lim_{t\rw\infty} D_{tv}\left( \Phi_{t|0}(\pi_0), \Phi_{t|0}(\tilde \pi_0) \right) = 0
$$ 
for any pair of prior probability measures $\pi_0, \tilde \pi_0 \in \mP(\mX)$.
\end{definition}

Very often stability is defined in a weaker form, by considering only prior measures $\pi_0$ and $\tilde \pi_0$ which are absolutely continuous w.r.t. each other. In this paper we refer to this property as {\em weak stability}.


The usual expression ``stability of the optimal filter'' may be misleading: As it is apparent from Definition \ref{defStability}, stability is a property of the operators $\Phi_{t|0}$, i.e., a property of the pair $\{\kappa, g \}$. As such, within this paper we often refer to the stability of the operators $\{\Phi_t\}_{t\ge 1}$ rather than the stability of the filters $\{ \pi_t \}_{t\ge 1}$.



\section{Truncated filters} \label{sTruncated}

%
\subsection{Truncation of state space models} \label{ssTruncation}

For an arbitrary but fixed sequence of observations $Y_{1:\infty}(\omega)=y_{1:\infty}$, let $\mS = \{ \pi_0, \kappa, g \}$ be a state space Markov model yielding the sequence of filters $\pi_t = \Phi_{t-1}(\pi_{t-1}) = g_t \cdot \kappa_t\pi_{t-1}$. We construct a truncated version of the model (and a sequence of filters for the truncated model) by
\begin{itemize}
\item[(i)] choosing a sequence of subsets of the state space $\mX$, denoted $\sfc := \{ C_t \}_{t \ge 1}$, where $C_t \subseteq \mX$ for $t\ge 1$
\item[(ii)] and defining the truncated potentials
\begin{equation} \label{eqTruncLikelihood}
g_t^\sfc(x) := \Ind_{C_t}(x) g_t(x),
\end{equation}
where $\Ind_{C_t}(x)$ is the indicator function, i.e., $g_t^\sfc(x)=g_t(x)$ for $x\in C_t$ and $g_t^\sfc(x)=0$ otherwise.
\end{itemize}
The truncated model is $\mS^\sfc = \{  \pi_0, \kappa, g^\sfc \}$, where $g^\sfc:=\{g_t^\sfc\}_{t\ge 1}$, and it yields the sequence of filters 
$$
\pi_t^\sfc = \Phi_t^\sfc(\pi_{t-1}) := g_t^\sfc \cdot \kappa_t \pi_{t-1}^\sfc
$$ 
and the sequence of predictive measures 
$$
\xi_t^\sfc = \Psi_t^\sfc(\pi_{t-1}) := \kappa_t\pi_{t-1}^\sfc,
$$ 
with $\pi_0^\sfc = \pi_0$ and composition operators denoted $\Phi_{t|0}^\sfc = \Phi_t^\sfc \circ\cdots\circ\Phi_1^\sfc$ and $\Psi_{t|0}^\sfc = \Psi_t^\sfc \circ\cdots\circ\Psi_1^\sfc$, respectively.

The truncated state space models constructed in this way have a simple but key feature: if one chooses a family of sets $\sfc^n = \{ C_t^n \}_{t \ge 1}$ such that \cblue{$C_t^n \subseteq C_t^{n+1}$ and $\lim_{n\rw\infty} C_t^n = \mX$ for every $t$ (meaning that $\lim_{n\rw\infty} \Ind_{C_t^n}(x) = 1$ when $x \in \mX$, and 0 otherwise),} then the sequence of truncated models $\mS^{\sfc,n}$ converges to the original model $\mS$ in the topology $\mD$ as $n\rw\infty$. This is made formal below.

\begin{lemma} \label{lmConvergence_in_D}
Let $\mS=\{\pi_0,\kappa,g\}$ be a state space model, let $\sfc^{n} = \{C_t^{n}\}_{t\ge 1}$ be a family of subsets of $\mX$ such that $\lim_{n\rw\infty} C_t^n = \mathcal X$ for every $t$, and denote $g_t^{\sfc,n} := \Ind_{C_t^n} g_t$. The sequence of truncated state space models $\mS^{\sfc,n}=\{\pi_0,\kappa, g^{\sfc,n}\}$, where $g^{\sfc,n}=\{ g_t^{\sfc,n} \}_{t\ge 1}$, converges to $\mS$ in the topology $\mD$. Moreover, if 
\begin{equation}
\lim_{n\rw\infty} \sup_{t\ge 1} \sup_{x \in \bar C^n_t}g_{t}(x)=0,
\label{eqAssThConv_in_D}
\end{equation}
then  $\lim_{n\rw\infty}D_{\mfM}(\mS^{\sfc,n}, \mS)=0$.
\end{lemma}
 
\begin{proof} 
Convergence in $\mD$ is trivial, since the prior $\pi_0$ and the kernel $\kappa_t$ is the same for all $n\ge 0$ and, clearly, $\lim_{n\rw\infty} g_t^{\sfc,n} = \lim_{n\rw\infty} g_t \Ind_{C_t^n} = g_t$ under the assumption $\lim_{n\rw\infty} C_t^n = \mX$. The definition of the metric $D_\mfM$ in \eqref{defMetric_in_S} together with assumption \eqref{eqAssThConv_in_D} readily yields $\lim_{n\rw\infty}D_{\mfM}(\mS^{\sfc,n}, \mS)=0$.
\end{proof}

%
\subsection{Stability of truncated PU operators}

Lemmas \ref{lmMarginals} and \ref{lmConvergence_in_D} together provide the means for the approximation of an arbitrary sequence of optimal filters $\pi_t$, generated by PU operators $\Phi_t$, by another sequence, $\pi_t^\sfc$, generated by truncated PU operators $\Phi_t^\sfc$. Unfortunately, truncation by itself does not guarantee that the new sequence of filters is stable. Below, we provide a stability theorem for sequences of truncated filters. 

\begin{theorem} \label{thStability}
Let $\sfc = \{ C_t \subseteq \mX\}_{t>0}$ be a sequence of subsets of the state space $\mX$ and let $\Phi_t^\sfc(\pi) := g_t^\sfc \cdot \kappa_t \pi$ be the truncated PU operator, where $\kappa_t$ is a Markov kernel, $g_t^\sfc = g_t\Ind_{C_t}$ is a truncated potential and $g_t$ is positive and bounded. If the Markov kernels $\kappa_t$ have positive pdf's $\sfk_t$ w.r.t. a reference probability measure $\lambda$, 
$$
\sfk_t(\cdot|x_{t-1})=\frac{\sd \kappa_t(\cdot | x_{t-1})}{\sd \lambda},
$$
such that
$$
\sum_{t=1}^{\infty} \frac{
        \inf_{ (x_{t-1}, x_t) \in C_{t-1} \times C_t} \sfk_t(x_t|x_{t-1})
}{
        \sup_{ (x_{t-1}, x_t) \in C_{t-1} \times C_t} \sfk_t(x_t|x_{t-1})
} =\infty,
$$
then the operator $\Phi_{t|0}^\sfc$ is stable, i.e.,
$$
\lim_{t\rw\infty}D_{tv}\left( \Phi_{t|0}^\sfc(\pi_0), \Phi_{t|0}^\sfc(\pi_0') \right) = 0
$$
for every $\pi_0, \pi_0' \in \mP(\mX)$ .
\end{theorem}

\begin{proof} 
See Lemma 3.1 in \cite{Crisan08}. The result is an extension of an original result in \cite{DelMoral01c} with methods introduced in \cite{Oudjane05}. \end{proof}

We are interested in truncated state space models $\mS^\sfc = \{ \pi_0, \kappa, g^\sfc \}$ that induce PU operators $\Phi_t^\sfc$ which can be proved to be stable. This is possible if the kernel $\kappa_t(\sd x|x')$ has a density $\sfk_t(x|x') \ge 0$ that satisfies the sufficient condition in Theorem \ref{thStability}, i.e., $\sum_{t=1}^\infty \varepsilon_t = \infty$, where 
$$
\varepsilon_t := \frac{
        \inf_{ (x_{t-1}, x_t) \in C_{t-1} \times C_t} \sfk_t(x_t|x_{t-1})
}{
        \sup_{ (x_{t-1}, x_t) \in C_{t-1} \times C_t} \sfk_t(x_t|x_{t-1})
}.
$$
In the next section we investigate a class of state space models and conditions on the choice of the subsets $\{C_t\}_{t\ge 1}$ for which the stability condition of Theorem \ref{thStability} can be guaranteed to hold.

\begin{remark}
An alternative to truncation for the construction of stable approximate filters is the iteration, at each time step, of a Markov kernel $M_{\pi_0,t}$ that leaves the filter $\pi_{t-1}=\Phi_{t-1|0}(\pi_0)$ invariant. To be specific, we can approximate the model $\mS=\{\pi_0,\kappa,g\}$ by another model $\hat \mS=\{\pi_0,\hat\kappa,g\}$, where $\hat \kappa=\{\hat \kappa_t\}_{t\ge 1}$ and $\hat \kappa_t=\kappa_t M_{\pi_0,t}^r$ for some integer $r \ge 1$. The PU operators for this model are denoted $\hat \Phi_t$. If the Dobrushin coefficient \cite{DelMoral01c} of the kernel $M_{\pi_0,t}$ is some $\beta_t < 1$, then,
by choosing $r$ large enough, one can ensure that the contraction due to the Markov kernel $M_{\pi_0,t}^r$ is sufficient to make the operator $\hat \Phi_{t|0}$ stable. Unfortunately, the kernel $M_{\pi_0,t}$ depends on $\kappa$, $g$ and the prior $\pi_0$ in a non-trivial manner and it is hard to compute it for most systems of interest.
\end{remark}

\section{Stable approximate filters} \label{sASSMs}

%
\subsection{A regular class of state space models} \label{ssAdditive}

In the sequel we study the class of state space models of the form $\mS=\{\pi_0,\kappa, g\}$ that satisfy the following regularity assumptions (recall that the potentials $g_t$ are positive and bounded real functions, i.e., $0<g_t<\| g_t \|_\infty<\infty$ for every $t\ge 1$).


\begin{assumption} \label{asLipschitz}
The conditional mean functions
\begin{equation}
a_t(x) := \int x' \kappa_t(\sd x'|x), \quad t \ge 1,
\nonumber
\end{equation}
are uniformly Lipschitz over time. To be specific, there exists $L_a<\infty$ such that
\begin{equation}
\sup_{t\ge 1} | a_t(x) - a_t(x') | < L_a \| x - x' \| \quad \mbox{for any $x,x'\in\mX$}.
\nonumber
\end{equation}
\end{assumption}

\begin{assumption} \label{asPDF}
The conditional probability measures $\kappa_t(\sd x| x')$ are absolutely continuous w.r.t. some probability measure $\lambda(\sd x)$ \cblue{with full support on $\mX$}, hence
$\kappa_t(\sd x | x') = \sfk_t(x|x')\lambda(\sd x),$
where $\sfk_t(x|x')$ is a conditional pdf. Moreover, there are \cblue{strictly} decreasing functions $\sfs_t : [0,\infty) \mapsto (0,\infty)$, $t \ge 1$, such that
\begin{equation}
\sfk_t(x|x') \ge \sfs_t(\|x-a_t(x')\|) > 0 \quad \mbox{and} \quad \lim_{r\rw\infty} \sfs_{\cblue{t}}(r)=0.
\nonumber
\end{equation} 
\end{assumption}

\begin{assumption} \label{asUpper}
The conditional pdf's $\sfk_t(x|x')$ are uniformly upper bounded over time. Specifically, there exists $C_0<\infty$  such that
\begin{equation}
\sup_{t\ge 1} \sup_{x,x'\in\mX} \sfk_t(x|x') < C_0.
\nonumber
\end{equation}
\end{assumption}

%

\subsection{Stable approximation via compact balls}

\label{ssCompactBalls}

In this section we show how it is possible to construct stable truncated approximations for state space models that satisfy Assumptions \ref{asLipschitz}--\ref{asUpper}. The specification of a truncated model relies on the choice of a sequence of subsets of the state space. Let us choose $\sfc^n=\{C_{t}^{n}\}_{t\geq 1}$, where $n$ is a positive integer and 
$$
C_{t}^{n} := B(\ell _{t},nr_{t}) = 
        \left\{ 
                x\in {\mathcal{X}} : \| x-\ell_{t} \| \leq nr_t
        \right\}, \quad t\geq 1,
$$
is the closed ball with centre $\ell _{t}\in {\mathcal{X}}$ and radius $nr_{t}>0$. The sequence $\{r_{t}\}_{t\geq 1}$ is selected to be positive and strictly increasing and
satisfy the identity 
\begin{equation}
\lim_{t\rw\infty }\frac{
        \sfs_{t}^{-1}\left( \upsilon_t \right) 
}{
        r_t
}=\infty,  \label{eqSlowEnough}
\end{equation}%
for the functions $\sfs_{t}$ in Assumption \ref{asPDF} and some strictly decreasing positive sequence $\{\upsilon_{t}\}_{t\geq 1}$ such that

\begin{equation}
\sum_{i\ge 1} \upsilon_{t_{i}}=\infty 
\label{eqDefUpsilon}
\end{equation}
for \cblue{any sequence  $\{t_i\}_{i \ge 1}$  such that $\liminf_{T\rw\infty} \frac{\left| \{ i \in \mathbb{N} : t_i<T \} \right|}{T} > 0.$}

\begin{remark}As we shall see below, condition \eqref{eqDefUpsilon} is the natural condition to impose in order to make the arguments work. In particular, if there exists a constant $c$ such that $v_n\ge \frac{c}{n}$ for all $n\ge 1$ the condition \eqref{eqDefUpsilon} is satisfied. See \cite{Lubeck18} for a proof of this result.    

\end{remark}

Intuitively, the time-dependent part of the radii, $r_t$, increases at a sufficiently slow rate compared to the sequence $\sfs_{t}^{-1}(\upsilon(t))$. 

The sequence of centres $\{\ell_{t}\}_{t\geq 1}$ is selected to satisfy the inequality
\begin{equation}
\| \ell_{t_i} - a_{\cblue{t_i}}(\ell_{{t_i}-1}) \| \le nLr_{t_i}  
\label{def_a_t}
\end{equation}%
for some constant $L<\infty$ and some sequence  $\{ t_i \}_{i \ge 1}$ such that
\begin{equation}
\liminf_{T\rw\infty} \frac{\left| \{ i \in \mathbb{N} : t_i<T \} \right|}{T} > 0.
\label{def_a_t_2}
\end{equation}


Given the family of sets $\sfc^n = \{ C_t^n \}_{t\ge 1}$ described above, we construct the truncated state space models $\mS^{\sfc,n} = \{ \pi_0, \kappa, g^{\sfc,n} \}$. The truncated PU operator is $\Phi_t^{\sfc,n}(\pi)=g_t^{\sfc,n}\cdot \kappa_t\pi$ and our aim is to prove that $\Phi_{t|0}^{\sfc,n}$ is stable for any integer $n$. Note that Lemma \ref{lmConvergence_in_D} ensures that $\lim_{n\rw\infty} \mS^{\sfc,n} = \mS$ in the topology ${\mathcal{D}}$ and, therefore, $\lim_{n\rw\infty} \Phi_{t|0}^{\sfc,n}(\pi_0) = \Phi_{t|0}(\pi_0) = \pi_t$ (via Lemma \ref{lmMarginals}).

The key result is stated and proved below. It yields a lower bound on the transition pdf between consecutive balls $C_{t_i-1}^n$ and $C_{t_i}^n$, where $\{t_i\}_{i\ge 1}$ is infinite sequence of time instants in the definition of the centres $\ell_t$ above (see Eqs. \eqref{def_a_t} and \eqref{def_a_t_2}).

\begin{lemma}
\label{lmInf_k} Choose any positive integer $n<\infty$ and let $\{t_i\}_{i\ge 1}$ be the infinite sequence in Eqs. \eqref{def_a_t} and \eqref{def_a_t_2}. If Assumptions \ref{asLipschitz} and \ref{asPDF} hold, then there exists $i_n<\infty$ such that 
$$
\inf_{(x,x') \in C_{t_i}^n \times C_{t_i-1}^n} \sfk_{t_i}(x|x') >  \upsilon_{t_i} \quad \mbox{for every $i>i_n$.} 
$$
\end{lemma}

\begin{proof} From Assumption \ref{asPDF} we have, 
\begin{equation}
\sfk_t(x|x^{\prime }) \ge \mathsf{s}_t(\|x-a_t(x')\|), \quad \forall x,x'\in {\mathcal{X}}.  \label{eqks}
\end{equation}
In particular, for any $x^{\prime }\in C_{t-1}^n = B(\ell_{t-1},nr_{t-1})$ and any $x \in C_t^n = B(\ell_t,nr_t)$, expression \eqref{eqks} together with a simple triangular inequality yields 
\begin{equation}
\sfk_t(x|x^{\prime }) \ge \mathsf{s}_t(\| x - \ell_t \| + \| \ell_t - a_t(\ell_{t-1}) \| + \| a_t(\ell_{t-1}) - a_t(x^{\prime }) \|),
\label{eqks1}
\end{equation}
where $\| x - \ell_t \| \le nr_t$ (since $x \in C_t^n$), and 
$$
\| a_t(\ell_{t-1}) - a_t(x') \| \le L_a \| \ell_{t-1} - x' \| \le L_a n r_{t-1},
$$
with $L_a < \infty$ independent of $t$, as a result of the Lipschitz Assumption \ref{asLipschitz} and the fact that $x'\in C_{t-1}^n$. Moreover, the choice of centres $\{ \ell_t \}_{t\ge 1}$ in \eqref{def_a_t} ensures that $\| \ell_{t_i} - a_{t_i}(\ell_{t_i-1}) \| \le n L r_{t_i}$ for the infinite sequence $\{t_i\}_{i\ge 1}$ in \eqref{def_a_t_2}.

Therefore, \eqref{eqks1} implies 
\begin{eqnarray}
\inf_{(x,x') \in C_{t_i}^n \times C_{t_i-1}^n} \sfk_{t_i}(x|x') &\ge& \sfs_{t_i}\left( n r_{t_i} + n L r_{t_i} + L_a n r_{t_i-1} \right)  \nonumber \\
&>& \sfs_{t_i}\left( n(1+ L + L_a)r_{t_i} \right)  \label{eqks3}
\end{eqnarray}
where the inequality \eqref{eqks3} holds because, by construction, $r_{t_i} > r_{t_i-1}$ and $\sfs_{t_i}$ is strictly decreasing.

However, the sequence $\{\cblue{r_t}\}_{t\geq 0}$ is chosen to increase ``slowly enough'' relative to the sequence $\sfs_{t}^{-1}(\cblue{\upsilon_t})$. Specifically, from the \cblue{identity} \eqref{eqSlowEnough} we deduce that there exists $i_n<\infty$ such that 
\begin{equation}
\sfs_{t_i}^{-1}(\cblue{\upsilon_{t_i}}) > n(1+L+L_{a})r_{t_i},
\quad \mbox{for every $i > i_n$,}  \label{eqks4}
\end{equation}
no matter the constants $L, L_{a}<\infty $. The inequalities \eqref{eqks3} and \eqref{eqks4} together imply that 
$$
\inf_{(x,x') \in C_{t_i}^{n} \times C_{t_i-1}^{n}} \sfk_{t_i}(x|x') >
\upsilon_{t_i},
\quad \forall i > i_n,
$$
which is obtained by applying the decreasing function $\sfs_{t_i}$ on both sides of \eqref{eqks4}.
\end{proof}


The stability of the truncated PU operators $\Phi_t^{\sfc,n}$ is a straightforward consequence of Lemma \ref{lmInf_k}.

\begin{theorem} \label{thStableSpheres} 
If Assumptions \ref{asLipschitz}--\ref{asUpper} hold then the PU operators $\Phi_t^{\sfc,n}$ are stable, i.e., 
$$
\lim_{t\rightarrow\infty} D_{tv}\left(  \Phi_{t|0}^{\sfc,n}(\pi_0), \Phi_{t|0}^{\sfc,n}(\pi_0') \right) = 0 
$$
for any $n<\infty$ and any pair of probability measures $\pi_0, \pi_0' \in {\mathcal{P}}({\mathcal{X}})$.
\end{theorem}

\begin{proof} Lemma \ref{lmInf_k} guarantees that there is some $i_n<\infty$ such that, for every $i > i_{n}$, we obtain 
\begin{equation}
\inf_{(x,x')\in C_{t_i}^{n} \times C_{t_i-1}^{n}} \sfk_{t_i}(x|x')> \upsilon_{t_i}.  
\nonumber 
\end{equation}%
Moreover, the latter inequality and Assumption \ref{asUpper} imply that, for all $i>i_n$, 
$$
\cblue{ \varepsilon_{t_i} > \frac{\upsilon_{t_i}}{C_0},  } \quad
\mbox{where} \quad
\varepsilon_t := \frac{
        \inf_{(x_{t-1},x_{t}) \in C_{t-1} \times C_{t}} \mathsf{k}_{t}(x_{t}|x_{t-1})
}{
        \sup_{(x_{t-1},x_{t})\in C_{t-1}\times C_{t}}\mathsf{k}_{t}(x_{t}|x_{t-1})
}. 
$$
However, Eq. \eqref{def_a_t_2} implies that \cblue{$\sum_{i > i_n} \upsilon_{t_i} = \infty$} which, together with Eq. \eqref{eqDefUpsilon}, yields
$$
\sum_{t\ge 1} \varepsilon_t \ge  \sum_{i > i_n} \varepsilon_{t_i} \ge \sum_{i > i_{n}} \upsilon_{t_i}=\infty 
$$
The inequality above ensures, via Theorem \ref{thStability}, that the operator $\Phi_{t|0}^{\sfc,n}$ is stable for any $n$.
\end{proof}


\begin{remark}
\label{rmFiniteHorizon} Let $\mfR\subset \mfM$ be the family of state space models that satisfy the regularity Assumptions \ref{asLipschitz}--\ref{asUpper}. Theorem \ref{thStableSpheres} implies that for any model $\mS \in \mfR$ it is possible to construct a sequence of truncated approximations $\mS^{\sfc,n}$, where the subsets in $\sfc=\{ C_t^n \}$ are closed balls of increasing radius, such that the associated PU operators $\Phi_t^{\sfc,n}$ are stable for every integer $n$.

Moreover, Lemma \ref{lmConvergence_in_D} ensures that $\lim_{n\rw\infty} {\mS}^{\sfc,n} = {\mathcal{S}}$ in the topology $\mD$. In particular, convergence in $\mD$ implies (via Lemma \ref{lmMarginals}) that $\lim_{n\rw\infty} \Phi_t^{\sfc,n}(\pi_0)=\Phi_t(\pi_0)$ for every $t$. Since $\pi_t^{\sfc,n} = \Phi_t^{\sfc,n}(\pi_0)$ is the filter at time $t$ generated by the truncated model $\mS^{\sfc,n}$, it follows that, for any finite time horizon $T<\infty$,
$$
\lim_{n\rw \infty}\max_{t\in [0,T]} D_{tv}( \pi_t^{\sfc,n}, \pi_t ) =0, 
$$
while guaranteeing that the sequence $\pi_t^{\sfc,n} = \Phi_{t|0}^{\sfc,n}(\pi_0)$ remains stable.
\end{remark}


\section{Example: stable approximation of an unstable filter} \label{sExample}

%
\subsection{State space model}

Let us consider the 1-dimensional, nonlinear state space model described by a prior $\pi_0 \in \mP(\Real)$ and the pair of equations
\begin{eqnarray}
X_t &=& s(X_{t-1})U_t, \label{eqStateEx}\\
Y_t &=& | X_t | + V_t, \label{eqObservationEx}
\end{eqnarray}
where $s(x)$ is the sign function\footnote{We define $s(x):=1$ for $x\ge 0$ and $s(x)=-1$ otherwise.}, $\{ U_t \}_{t\ge1}$ is a sequence of truncated normal r.v.'s, namely $U_t \sim \mathcal{TN}\left(|X_{t-1}|,\sigma_u^2,[0,+\infty)\right)$, and $\{V_t\}_{t \ge 1}$ is an i.i.d. sequence of normal r.v.'s, namely $V_t \sim \mN(0,\sigma_v^2)$.

Let $F_N(x)$ and $f_N(x)$ denote the cumulative distribution function (cdf) and the pdf, respectively, of the standard normal distribution, $\mN(0,1)$. The Markov kernel $\kappa_t(\sd x | x')$ for this model has a pdf w.r.t. Lebesgue measure that can be explicitly written as
\begin{equation}
\sfk_t(x|x') = \frac{
        \exp\left\{
                \frac{1}{-2\sigma_u^2} \left( x - x' \right)^2
        \right\} \left(
                \Ind_{(-\infty,0)^2}(x,x')) + \Ind_{(0,\infty)^2}(x,x')
        \right)
}{
        F_N\left( \frac{|x'|}{\sigma_u} \right) \sqrt{ 2\pi\sigma_u^2 } 
} \label{eqTransPDF}
\end{equation}
and the mean function can be shown to yield
\begin{equation}
a_t(x') = \int x \sfk_t(x|x') \sd x' = x' + s(x')\frac{
                f_N\left( \frac{|x'|}{\sigma_u} \right)
        }{
                F_N\left(  \frac{|x'|}{\sigma_u} \right)
        } \sigma_u.
%
\label{eqMeanFunctionEx}
\end{equation}
As the observation noise $V_t$ is Gaussian, the potential function has the form
$$
g_t(x_t) = \frac{1}{\sqrt{2\pi\sigma_v^2}} \exp\left\{
        -\frac{1}{2\sigma_v^2} (y_t-|x_t|)^2
\right\}.
$$

A key feature of model \eqref{eqStateEx}-\eqref{eqObservationEx} is that $X_tX_{t-1}>0$ for every $t$, i.e., the sequence of states $X_{1:\infty}=x_{1:\infty}$ is either all-positive or all-negative. Given this property, it is natural to decompose the prior measure $\pi_0$ into positive and negative parts, namely
\begin{equation}
\pi_0 = \pi_0\left((0,\infty)\right) \pi_0^+ + \pi_0\left((-\infty,0)\right) \pi_0^-,
\label{eq4app-0}
\end{equation}
where we define the probability measures $\pi_0^+$ and $\pi_0^-$ as
$$
\pi_0^+(A) := \frac{
        \pi_0\left(A\cap(0,\infty)\right)
}{
        \pi_0\left((0,\infty)\right)
} \quad \mbox{and} \quad
\pi_0^-(A) :=  \frac{
        \pi_0\left( A\cap (-\infty, 0) \right)
}{
        \pi_0\left( (-\infty,0) \right)
},
$$
respectively, for any Borel subset $A \subset \Real$. This decomposition can be ``propagated'' to time $t>0$ as stated in Proposition \ref{propDecompose} below. In order to state this result, let us denote
\begin{equation}
\ell_{0,\pi_0}^+=\pi_0\left((0,\infty)\right) \quad \mbox{and}\quad
\ell_{0,\pi_0}^-=\pi_0\left((-\infty,0)\right)
\label{eqEll0}
\end{equation}
and let $\Phi_t(\alpha)=g_t\cdot\kappa_t\alpha$ be the PU operator generated by model \eqref{eqStateEx}--\eqref{eqObservationEx}.

\begin{proposition} \label{propDecompose}
If $\pi_0^+$ and $\pi_0^-$ are both non-null, then the optimal filter at time $t$ can be decomposed as 
\begin{equation}
\pi_t = \ell_{t,\pi_0}^+ \pi_t^+ + \ell_{t,\pi_0}^- \pi_t^-, \label{eqDecompo_t}
\end{equation}
where $\pi_t^+ = \Phi_{t|0}(\pi_0^+)$ and $\pi_t^- = \Phi_{t|0}(\pi_0^-)$ are probability measures, and the linear combination coefficients $\ell_{t,\pi_0}^+$ and  $\ell_{t,\pi_0}^-$ are constructed recursively as
\begin{eqnarray}
\ell_{t,\pi_0}^+ = \ell_{t-1,\pi_0}^+ \frac{
        \left( g_t, \kappa_t\Phi_{t-1|0}(\pi_0^+) \right)
}{
        \left( g_t, \kappa_t\Phi_{t-1|0}(\pi_0) \right)
}, \label{eqRecurs1}\\
\ell_{t,\pi_0}^- = \ell_{t-1,\pi_0}^- \frac{
        \left( g_t, \kappa_t\Phi_{t-1|0}(\pi_0^-) \right)
}{
        \left( g_t, \kappa_t\Phi_{t-1|0}(\pi_0) \right)
}, \label{eqRecurs2}
\end{eqnarray}
for $t \ge 1$. 
\end{proposition}

\begin{proof} 
See Appendix \ref{apDecompose}.
\end{proof}

The PU operator $\Phi_t$ for model  \eqref{eqStateEx}--\eqref{eqObservationEx} can be unstable. To see, this, let $\Phi_t^{Y_t}(\alpha)=g_t^{Y_t} \cdot \kappa_t\alpha$ be the random PU operator induced by the r.v. $Y_t$. We can now introduce the set
$$
A_u := \left\{
        \omega \in \Omega : \quad \{ \Phi_t^{Y_t(\omega)} \}_{t \ge 1} \quad \mbox{is unstable} \right\},
$$
which describes all possible realisations of the observations process $Y$ that yield an unstable sequence of filters. Proposition \ref{propUnstable} below states that the probability of this set is positive.

\begin{proposition} \label{propUnstable}
The set $A_u$ corresponding to the state space model described by the equations \eqref{eqStateEx}--\eqref{eqObservationEx} is non-negligible, i.e., $\Prob(A_u)>0.$ 
\end{proposition}

\begin{proof}  
See Appendix \ref{apUnstable}.
\end{proof}

%
\subsection{Stable truncated approximation}

The model \eqref{eqStateEx}--\eqref{eqObservationEx} can yield unstable PU operators $\Phi_t$. However,  we can still construct a stable approximation by truncation of the positive and negative parts of $\pi_t$. In particular, from Eq. \eqref{eqTransPDF} we readily obtain the conditional pdf's
\begin{eqnarray}
\sfk_t^+(x|x') &=& \frac{
        1
}{
        F_N\left( \frac{|x'|}{\sigma_u} \right) \sqrt{ 2\pi\sigma_u^2 } 
} \exp\left\{
        -\frac{ \left( x - x' \right)^2}{2\sigma_u^2}
\right\} \Ind_{(0,\infty)^2}(x,x') \nonumber\\
\sfk_t^-(x|x') &=& \frac{
        1
}{
        F_N\left( \frac{|x'|}{\sigma_u} \right) \sqrt{ 2\pi\sigma_u^2 } 
} \exp\left\{
        -\frac{\left( x - x' \right)^2}{2\sigma_u^2} 
\right\} \Ind_{(-\infty,0)^2}(x,x') \nonumber
\end{eqnarray}
and the potentials
$$
g_t^+(x) = \frac{1}{\sqrt{2\pi\sigma_v^2}} \exp\left\{
        -\frac{ (y_t-x)^2}{2\sigma_v^2}
\right\}, \quad
g_t^-(x) = \frac{1}{\sqrt{2\pi\sigma_v^2}} \exp\left\{
        -\frac{(y_t+x)^2}{2\sigma_v^2} 
\right\} 
$$
in such a way that
$$
\pi_t^+ = \Phi_{t|0}^+(\pi_0^+) = \Phi_{t|0}(\pi_0^+)
\quad \mbox{and}\quad
\pi_t^- = \Phi_{t|0}^-(\pi_0^-) = \Phi_{t|0}(\pi_0^-),
$$
where $\Phi_t^+(\alpha) = g_t^+\cdot \kappa_t^+\alpha$ and $\Phi_{t|0}^-(\alpha) = g_t^- \cdot \kappa_t^- \alpha$. It is apparent that
\begin{itemize}
\item both $g_t^+$ and $g_t^-$ are positive and bounded, hence they satisfy Assumption \ref{asLipschitz};
\item both $\sfk_t^+(x|x')$ and $\sfk_t^-(x|x')$ are uniformly upper bounded and so satisfy Assumption \ref{asUpper}.
\end{itemize}
The mean functions for the kernels $\sfk_t^+$ and $\sfk_t^-$ both have the form in \eqref{eqMeanFunctionEx}, i.e., $a_t^+(x)=a_t(x)\Ind_{(0,\infty)}(x)$ and $a_t^-(x)=a_t(x)\Ind_{(-\infty,0)}(x)$. Both functions are Lipschitz and, therefore, satisfy Assumption \ref{asLipschitz}. To see this, let us consider the case of $a_t^+(x)$. Since $x>0$, the derivative w.r.t. $x$ can be calculated exactly and it yields
$$
\frac{\sd a_t^+}{\sd x}(x) = 1 + \frac{
                f_N'\left( \frac{x}{\sigma_u} \right) F_N\left(  \frac{x}{\sigma_u} \right) 
                - f_N\left( \frac{x}{\sigma_u} \right) F_N'\left(  \frac{x}{\sigma_u} \right)
        }{
                \left( F_N\left(  \frac{x}{\sigma_u} \right) \right)^2
        }, \quad \mbox{for $x>0$},
$$ 
where $f_N'=\frac{\sd f_N}{\sd x}$ and $F_N'=\frac{\sd F_N}{\sd x}$. However,
\begin{itemize}
\item both $f_N$ and $F_N$ are Lipschitz, hence there exist $C_f<\infty$ and $C_F<\infty$ such that $|f_N'|<C_f$ and $|F_N'|<C_F$, respectively, and
\item $F_N(x) \ge \frac{1}{2}$ for $x>0$.
\end{itemize}
Therefore, recalling that $f_N \le  \left( 2\pi \right)^{-\frac{1}{2}}$ and $F_N\le 1$,
$$
\left|
        \frac{\sd a_t^+}{\sd x}(x)
\right| < 1 + 4 \left(
        C_f + \left( 2\pi \right)^{-\frac{1}{2}} C_F
\right) < \infty,
$$
i.e., $a_t^+(x)$ is Lipschitz. The calculations for $a_t^-$ are similar.

As for Assumption \ref{asPDF}, since 
$$
|x'-a_t^+(x')| < \frac{2}{\sqrt{2\pi\sigma_u^2}}, 
$$
we can always choose a sufficiently small constant $c_4>0$ such that
$$
\sfk_t^+(x|x') > \sfs_t\left( |x-a_t^+(x')| \right) := c_4
\exp\left\{
        -\frac{1}{2\sigma_u^2} \left( x - a_t^+(x') \right)^4
\right\}, \quad \mbox{for $x,x'>0$},
$$
and, similarly, $\sfk_t^-(x|x') > \sfs_t\left( |x-a_t^-(x')| \right)$ for $x,x'<0$. Therefore, Assumption \ref{asPDF} holds both for $\Phi_t^+$ and $\Phi_t^-$.

Finally, since the PU operators $\Phi_t^+$ and $\Phi_t^-$ correspond to state space models in $\mfR$, we can construct truncated stable approximations $\pi_t^{\sfc,n,+}$ and $\pi_t^{\sfc,n,-}$ using the method in Section \ref{ssCompactBalls} (see Theorem \ref{thStableSpheres}). In particular, we construct the balls $C_t^n = B(\ell_t, nr_t)$, where $\ell_t^n=y_t$ and $r_t$ is any increasing sequence that satisfies 
$
\lim_{t\rw\infty} r_t^{-1} \sfs_t^{-1}(\upsilon_t) = \infty.
$
The choice of $\ell_t = y_t$ guarantees that the inequalities \eqref{def_a_t} and \eqref{def_a_t_2} hold, as stated by Proposition \ref{propCentres} below.

\begin{proposition} \label{propCentres}
Consider the state space models $\mS^+ = \{ \pi_0^+, \kappa^+, g^{Y,+} \}$, where $g^{Y,+} = \{ \Ind_{(0,\infty)}g_t^{Y_t} \}_{t\ge 1}$. If we let $\ell_t=Y_t$, $t \ge 1$, then there exists a.s. an infinite sequence $\{ t_i \}_{i \ge 1}$ such that 
\begin{equation}
\left| Y_{t_i} - a_{t_i}^+(Y_{t_i-1}) \right| < L r_t \quad 
\mbox{and} \quad
\lim_{T\rw\infty} \frac{
        \left| \{ i: t_i < T \} \right|
}{
        T
} > 0,
\label{eqTargetStatement}
\end{equation}
for some constant $L<\infty$.
\end{proposition}
\begin{proof} A simple triangle inequality yields
\begin{eqnarray}
\left| Y_t - a_t^+(Y_{t-1}) \right| &\le& \left|
        Y_t - a_t^+(X_{t-1})
\right| +  \left|
        a_t^+(X_{t-1}) - a_t^+(Y_{t-1}) 
\right| \nonumber\\
&\le& |V_t| + L_a|V_{t-1}| + |X_t -a_t^+(X_{t-1})|, \nonumber
\end{eqnarray}
where the second inequality is obtained by recalling that $Y_t=X_t+V_t$ and $L_a$ is the Lipschitz constant of function $a_t^+$. Moreover, since
$$
a_t^+(x) = x + \sigma_u\frac{
        f_N\left(\frac{x}{\sigma_u}\right)
}{
        F_N\left(\frac{x}{\sigma_u}\right)
} \le x + \sqrt{\frac{2\sigma_u^2}{\pi}},
$$
we readily arrive at 
\begin{equation}
\left| Y_t - a_t^+(Y_{t-1}) \right| \le |V_t| + L_a|V_{t-1}| + |Z_t| + c,
\label{eqRoughBound}
\end{equation}
where $Z_t := X_t-X_{t-1} \sim \mathcal{TN}\left(0,\sigma_u^2,(-X_{t-1},\infty)\right)$ and $c=\sqrt{\frac{2\sigma_u^2}{\pi}}$. From \eqref{eqRoughBound} we deduce that
\begin{eqnarray}
\Prob\left(
        \left| Y_t - a_t^+(Y_{t-1}) \right| > r_t
\right) &\le& \Prob\left(
        |V_t| + L_a|V_{t-1}| + |Z_t| + c > r_t
\right) \nonumber \\
&\le& \Prob\left(
        |V_t| + L_a|V_{t-1}| + |Z| > r_t
\right), \label{eqSoftBound}
\end{eqnarray}
where $Z \sim \mN(c,\sigma_u^2)$, and, since $V_t$ is an i.i.d. sequence and $Z$ is independent of $t$, the fact that $\lim_{t\rw\infty} r_t=\infty$ implies that 
\begin{equation}
\lim_{t\rw\infty} \Prob\left(
        \left| Y_t - a_t^+(Y_{t-1}) \right| > r_t
\right) \le \lim_{t\rw\infty} \Prob\left(
        |V_t| + L_a|V_{t-1}| + |Z| > r_t
\right) = 0.
\label{eqConvProb}
\end{equation}

The limit in \eqref{eqConvProb} implies that for any $\epsilon>0$ we can find $L_\epsilon < \infty$ such that 
$$
\Prob\left(
        \left| Y_t - a_t^+(Y_{t-1}) \right| > L_\epsilon r_t
\right) < \epsilon
$$ 
for every $t$. Then, L\'evy's extension of the Borel-Cantelli lemmas \cite{Williams91} implies \eqref{eqTargetStatement}. To be precise, there is an infinite sequence $\{ t_i \}_{i\ge 1}$ such that $\left| Y_{t_i} - a_{t_i}^+(Y_{t_i-1}) \right| < L_\epsilon r_{t_i}$ and $\lim_{T\rw\infty} T^{-1} |\{t_i\}_{i\ge 1}| > 1-\epsilon$.
\end{proof}

We can prove the same result for $\mS^- = \{ \pi_0^-, \kappa^-, g^{Y,-} \}$ with the same argument.

Finally, since $\mS^+$ and $\mS^-$ are in the class $\mfR$, the truncated filters $\pi_t^{\sfc,n,+}$ and $\pi_t^{\sfc,n,-}$ are stable for every $n$, and it readily follows that 
$
\pi_t^{\sfc,n} = \ell_{t,\pi_0}^+ \pi_t^{\sfc,n,+} + \ell_{t,\pi_0}^- \pi_t^{\sfc,n,-}
$ 
is also stable. Moreover, since $\lim_{n\rw\infty} \pi_t^{\sfc,n,+} = \pi_t^+$ and $\lim_{n\rw\infty} \pi_t^{\sfc,n,-} = \pi_t^-$, we arrive at
$$
\lim_{n\rw\infty}  \pi_t^{\sfc,n} =  
\ell_{t,\pi_0}^+ \lim_{n\rw\infty}  \pi_t^{\sfc,n,+} + \ell_{t,\pi_0}^- \lim_{n\rw\infty}  \pi_t^{\sfc,n,-} = 
\ell_{t,\pi_0}^+ \pi_t^+ + \ell_{t,\pi_0}^- \pi_t^- = \pi_t.
$$

%
\subsection{Extension to higher dimensional spaces}

It is straightforward to extend model \eqref{eqStateEx}--\eqref{eqObservationEx} to a general class of state space models on a $d_x$-dimensional space $\mX \subseteq \Real^{d_x}$. In particular, let us choose a partition 
$$
\mX = \bigcup_{i=1}^n A_i, \quad \mbox{where} \quad A_i \cap A_j = \emptyset \quad \mbox{whenever $i\ne j$.}
$$
Assume that, for any subset $B \in \mB(\mX)$, $\kappa_t(B|x\in A_i)=0,$ if $\cap A_i = \emptyset$ i.e., once the state is contained in $A_i$, it remains in that subset. Then, it is straightforward to construct measures $\pi_0^1, \ldots, \pi_0^n$ such that $\pi_0^i(A_i)=1$ and $\pi_0 = \sum_{i=1}^n \ell_{0}^i\pi_0^i$. Moreover, the optimal filter at time $t$ can be expressed as
$$
\pi_t = \sum_{i=1}^n \ell_{t,\pi_0}^i \pi_t^i,
$$
where $\pi_t^i = \Phi_{t|0}(\pi_0^i)$ and
$
\ell_{t,\pi_0}^i = \ell_{t-1,\pi_0}^i \frac{
        (g_t,\kappa_t\Phi_{t|0}(\pi_0^i))
}{
        (g_t,\kappa_t\Phi_{t|0}(\pi_0))
}.
$

\section{Uniform approximation over time} \label{sUniform}

In this section we explore alternative approximations of the state space model $\mS=\{\pi_0,\kappa,g\}$ where, besides truncating the likelihoods $g_t$, we {\em reshape} the Markov kernels $\kappa_t$ in accordance with the sequence of sets $\sfc=\{C_t\}_{t\ge 1}$. These modified kernels can be obtained from any given $\kappa_t$ as described below.

\cblue{
\begin{definition}
\label{defReshaped} Let $\mS=\{\pi_0,\kappa,g\}$ be a state space model and let $\sfc = \{C_t\}_{t\ge 1}$ be a sequence of subsets of ${\mathcal{X}}$. We define the ``reshaped'' Markov kernel $\tilde \kappa^\sfc_t$ as
\begin{equation}
\tilde \kappa^\sfc_1 := \kappa_1 \quad \mbox{and} \quad \tilde \kappa^\sfc_t(\sd x|x') := \kappa_t(\sd x | x') \pi_{t-1}(C_{t-1}) + \rho_t(\sd x)  \label{eqDefKappa2}
\end{equation}
for $t \ge 2$, where 
\begin{equation}
\rho_t(\sd x) := \int \mathbbm{1}_{\bar C_{t-1}}(x') \kappa_t(\sd x | x') \pi_{t-1}(\sd x') =
\left( \kappa_t\Ind_{\bar C_{t-1}}, \pi_{t-1} \right)(\sd x).
\label{eqDefRho}
\end{equation}
%
%
\end{definition}
}

We now investigate the use of reshaped kernels to build approximate state space models and sequences of optimal filters. First we identify \cblue{a class of truncated filters that employ the reshaped Markov kernels introduced in Definition \ref{defReshaped} and establish its key properties.} Then, we prove that, provided each set in the sequence $\sfc = \{ C_t \}_{t \ge 0}$ contains a sufficiently large probability mass, the truncated filters can be kept arbitrarily close to the optimal filters uniformly over time.

\cblue{
\begin{lemma}
\label{lmReshaped} Let $\mS=\{\pi_0,\kappa,g\}$ be a state space model and let $\sfc = \{C_t\}_{t\ge 1}$ be a sequence of subsets of $\mX$. The truncated state space model $\tilde \mS^\sfc=\{\pi_0,\tilde \kappa^\sfc, g^\sfc\}$, where $\tilde \kappa_t^\sfc = \{ \tilde \kappa_t^\sfc \}_{t\ge 1}$ and $g^\sfc = \{g_t^\sfc\}_{t\ge 1}$, yields sequences of predictive and filtering probability measures ($\tilde \xi_t^\sfc$ and $\tilde \pi_t^\sfc$, respectively) such that, for any integrable $f:\mX\mapsto\Real$ and every $t \ge 1$, 
\begin{eqnarray}
(\Ind_{C_t} f, \xi_t) &=& (\Ind_{C_t} f, \tilde \xi_t^\sfc), \quad \mbox{and}  \label{eqXi-Ct} \\
(\Ind_{C_t} f, \pi_t) &=& (f, \tilde \pi_t^\sfc) \pi_t(C_t).
\label{eqPi-Ct}
\end{eqnarray}
%
\end{lemma}
}

\begin{proof} 
See Appendix \ref{apReshaped}.
\end{proof}


\cblue{
\begin{remark}
The identity \eqref{eqPi-Ct} shows that the truncated filter $\tilde \pi_t^\sfc$ can be seen as the restriction of the original filter $\pi_t$ to the set $C_t$, with the normalisation constant $\pi_t(C_t)$ needed to ensure that $\tilde \pi_t^\sfc$ is a probability measure.
\end{remark}
}

From Lemma \ref{lmReshaped} it is relatively easy to show that the approximation error $D_{tv}(\pi_t,\tilde \pi_t^\sfc)$ can be uniformly controlled over time.

\begin{theorem}
\label{thEpsilonError} Let ${\mathcal{S}}=\{\pi_0,\kappa_t,g_t\}$ be a state space model and let $\sfc = \{C_t\}_{t\ge 1}$ be a sequence of subsets of $\mX$. Assume every $C_t$ is large enough to ensure that 
\begin{equation}
\pi_t(\bar C_t)=(\Ind_{\bar C_t},\pi_t) < \cblue{\frac{1}{2}} \epsilon
\label{eqAssumptionEps}
\end{equation}
for some prescribed $0<\epsilon<1$. Then, the truncated state space model $\tilde \mS^\sfc=\{\pi_0,\tilde \kappa^\sfc, g^\sfc\}$ yields a sequence of filters $\tilde \pi_t^\sfc = \tilde \Phi_{t|0}^\sfc(\pi_0)$ such that $\sup_{t\ge 0}D_{tv}(\pi_t, \tilde \pi_t^\sfc) < \epsilon$. 
\end{theorem}

\begin{proof} 
\cblue{
Let $\tilde \xi_t^\sfc = \tilde \kappa_t^\sfc \tilde \pi_{t-1}^\sfc$ be the sequence of predictive probability measures generated by the truncated model $\tilde \mS^\sfc$. Using the relationship \eqref{eqOperators2} and the definition $g_t^\sfc=\Ind_{C_t}g_t$, the approximation error can be written as 
\begin{eqnarray}
| (f,\pi_t) - (f,\tilde\pi_t^\sfc) | &=& \left|
        \frac{
                (fg_t, \xi_t)
        }{
                (g_t,\xi_t)
        } \pm \frac{
                (f\Ind_{C_t}g_t, \tilde\xi_t^\sfc)
        }{
                (g_t,\xi_t)
        }
- \frac{
                (f\Ind_{C_t}g_t, \tilde\xi_t^\sfc)
        }{
                (\Ind_{C_t}g_t,\tilde\xi_t^\sfc)
        }
\right| \nonumber \\
&\le& \left|
        \frac{
                (fg_t, \xi_t)
        }{
                (g_t,\xi_t)
        } - \frac{
                (f\Ind_{C_t}g_t, \tilde\xi_t^\sfc)
        }{
                (g_t,\xi_t)
        } 
\right| + \left|
        \frac{
                (f\Ind_{C_t}g_t, \tilde\xi_t^\sfc)
        }{
                (g_t,\xi_t)
        } - \frac{
                (f\Ind_{C_t}g_t, \tilde\xi_t^\sfc)
        }{
                (\Ind_{C_t}g_t,\tilde\xi_t^\sfc)
        }
\right|. \label{eqIneq0}
\end{eqnarray}
Let us consider the first term on the right-hand side of \eqref{eqIneq0}. From Lemma \ref{lmReshaped}, we have $(f\Ind_{C_t}g_t, \tilde \xi_t^\sfc) = (f\Ind_{C_t}g_t, \xi_t)$ and, therefore,
\begin{eqnarray}
\left|
        \frac{
                (fg_t, \xi_t)
        }{
                (g_t,\xi_t)
        } - \frac{
                (f\Ind_{C_t}g_t,\tilde \xi_t^\sfc)
        }{
                (g_t,\xi_t)
        } 
\right| &=& \left|
        \frac{
                (fg_t, \xi_t)
        }{
                (g_t,\xi_t)
        } - \frac{
                (f\Ind_{C_t}g_t, \xi_t)
        }{
                (g_t,\xi_t)
        } 
\right| \nonumber\\
&=& \left|
        (f,\pi_t) - (\Ind_{C_t} f, \pi_t)
\right| \nonumber \\
&=& | (\Ind_{\bar C_t} f, \pi_t) | 
\le  \frac{1}{2} \| f \|_\infty \epsilon,
\label{eqIneq1}
\end{eqnarray}
where the inequality follows from \eqref{eqAssumptionEps}.
}

\cblue{
We can rewrite the second term on the right-hand side of \eqref{eqIneq0} as
\begin{eqnarray}
\left|
        \frac{
                (f\Ind_{C_t}g_t, \tilde \xi_t^\sfc)
        }{
                (g_t,\xi_t)
        } - \frac{
                (f\Ind_{C_t}g_t, \tilde \xi_t^\sfc)
        }{
                (\Ind_{C_t}g_t,\tilde \xi_t^\sfc)
        }
\right| &=& \left|
        \frac{
                (f\Ind_{C_t}g_t, \tilde \xi_t^\sfc)
        }{
                (\Ind_{C_t}g_t,\tilde \xi_t^\sfc)
        } \times \frac{
                (\Ind_{C_t}g_t,\tilde \xi_t^\sfc) - (g_t,\xi_t)
        }{
                (g_t,\xi_t)
        }
\right| \nonumber \\
&=& \left| (f,\tilde \pi_t^\sfc) \right| \times \left|
        \frac{
                (\Ind_{\bar C_t}g_t,\xi_t)
        }{
                (g_t,\xi_t)
        }
\right| \le \frac{1}{2} \| f \|_\infty \epsilon, \label{eqNearly}
\end{eqnarray}
where the second equality follows from the relationship $\tilde \pi_t^\sfc = g_t^\sfc \cdot  \tilde\xi_t^\sfc$ and Lemma \ref{lmReshaped} and the final inequality is a consequence of  \eqref{eqAssumptionEps}. 
}

\cblue{
Substituting \eqref{eqIneq1} and \eqref{eqNearly} into \eqref{eqIneq0} yields $| (f,\pi_t) - (f,\tilde\pi_t^\sfc) | \le \| f \|_\infty \epsilon$ and, therefore,
$
D_{tv}\left( \pi_t, \tilde \pi_t^\sfc \right) = \sup_{\{f\in \mathsf{B}(\mathcal{X}), 0\le f\le 1\}}| (f,\pi_t) - (f, \tilde \pi_t^\sfc)|\le \epsilon.
$
}
\end{proof}

For the family of sets $\sfc^n=\{C_t^n\}$, let $\tilde \kappa^{\sfc,n}_t$ be the reshaped kernel constructed as in Definition \ref{defReshaped}(ii) for the set $C_t^n$ and let $g_t^{\sfc,n}=\Ind_{C_t^n}g_t$. The approximate models $\tilde \mS^{\sfc,n}=\{\pi_0,\tilde \kappa^{\sfc,n},g^{\sfc,n}\}$, where $\tilde \kappa^{\sfc,n} = \{ \kappa_t^{\sfc,n} \}_{t\ge 1}$ and $g^{\sfc,n} = \{ g_t^{\sfc,n} \}_{t\ge 1}$, preserve convergence in the topology $\mD$, in a way similar to the original truncated models $\mS^{\sfc,n}$. In particular, we have the following result.

\begin{theorem}
\label{thConvergence_in_D2} Let ${\mathcal{S}}=\{\pi_0,\kappa,g\}$ be a state space model and let $\sfc^{n} = \{C_t^{n}\}_{t\ge 1}$ be a sequence of subsets of ${\mathcal{X}}$ such that $\lim_{n\rightarrow\infty} C_t^n = \mathcal{X}$. The sequence of truncated models $\tilde \mS^{\sfc,n}=\{\pi_0,\tilde \kappa^{\sfc,n}, g^{\sfc,n} \}$ converges to $\mS$ in the topology $\mD$. Moreover, if 
\begin{equation}
\lim_{n\rightarrow\infty}\sup_{t\ge 1} \pi_t(\bar C_{t}^n) =
\lim_{n\rightarrow\infty} \sup_{t\ge 1} \sup_{x \in \bar C^n_t}g_{t}(x)=0,
\label{eqAssThConv_in_D2}
\end{equation}
then $\lim_{n\rightarrow\infty}D_\mfM(\tilde \mS^{\sfc,n}, \mS)=0.$
\end{theorem}

\begin{proof}
As the initial condition $\pi_0$ is the same for all $n\ge 0$ and, trivially, $\lim_{n\rightarrow\infty} g_t^{\sfc,n} = \lim_{n\rightarrow\infty} g_t \mathbbm{1}_{C_t^n} = g_t$ under the assumption $\lim_{n\rightarrow\infty} C_t^n = {\mathcal{X}}$, it suffices to show that 
\begin{equation}
\lim_{n\rightarrow\infty} D_{tv}( \tilde \kappa^{\sfc,n}_t(\cdot,x), \kappa_t(\cdot,x) ) = 0  \label{eqEnough2}
\end{equation}
for any $x\in\mathcal{X}$. However, from Definition \ref{defReshaped} we readily obtain that 
\begin{eqnarray*}
\left| \left( f, \tilde \kappa^{\sfc,n}_t(\cdot,x)) - (f,\kappa_t(\cdot,x) \right) \right| &\cblue{\le}& 
\left( f,\kappa_t(\cdot,x) \right) \pi_{t-1}(\bar C_{t-1}^n) + \left( (f,\kappa_t) \mathbbm{1}_{\bar C_{t-1}^n}, \pi_{t-1} \right) \\
&\le& 2 \| f \|_{\infty} \pi_{t-1}(\bar C_{t-1}^n),
\end{eqnarray*}
where $\lim_{n\rightarrow\infty} \pi_{t-1}(\bar C_{t-1}^n) = 0$ (since $\lim_{n\rw \infty} C_{t-1}^n = \cblue{\mX}$), hence \eqref{eqEnough2} is satisfied.

For the second claim, we deduce from the above that 
\begin{equation}
D_\mfM(\tilde \mS^{\sfc,n}, \mS ) = 2\sup_{t\ge 1}\pi_{t-1}(\bar C_{t-1}^n) + \sup_{t\ge 1}\sup_{x\in \bar C^n_t}g_{t}(x),  \label{eqDistance2}
\end{equation}
which implies $\lim_{n\rightarrow\infty} D_\mfM(\tilde \mS^{\sfc,n}, \mS)=0$ as a consequence of the assumption \eqref{eqAssThConv_in_D2}.
\end{proof}

Let $\mS$ be a state space model of the class $\mfR$ introduced in Remark \ref{rmFiniteHorizon} and construct the sets in the family $\sfc^n=\{C_t^n\}_{t\ge 1}$ in the same way as in Section \ref{ssCompactBalls}, i.e., $C_t^n = B(\ell_t,nr_t)$. The operators $\tilde \Phi_{t|0}^{\sfc,n}$ associated to the truncated model $\tilde \mS^{\sfc,n}$ can be proved to be stable for every integer $n$ provided that $\inf_{t\ge 1} \pi_t(C_t^n)>\epsilon>0$.


\begin{theorem}
\label{thStableSpheres-2} 
Let $\mS=\{\pi_0,\kappa,g\}$ be a state space model of the class $\mfR$ and assume that there exists $\epsilon>0$ such that 
\begin{equation}
\inf_{t \ge 1} \pi_t(C_t^n) \ge \epsilon.  \label{eqBigenough}
\end{equation}
Then the PU operators $\tilde \Phi_t^{\sfc,n}$, $t\ge 1$, are stable, i.e., 
\begin{equation*}
\lim_{t\rightarrow\infty} D_{tv}\left(  \tilde \Phi_{t|0}^{\sfc,n}(\pi), \Phi_{t|0}^{\sfc,n}(\pi') \right) = 0 
\end{equation*}
for any $n<\infty$ and any pair of probability measures $\pi, \pi' \in \mP(\mX)$.
\end{theorem}


\begin{proof}
Lemma \ref{lmInf_k} guarantees that there is some $i_n<\infty$ such that, for every $i>i_n$,
\begin{equation}
\inf_{(x,x') \in C_{t_i}^n \times C_{t_i-1}^n} \sfk_{t_i}(x|x') > \upsilon_{t_i},  \label{eqSph0.0-2}
\end{equation}
where $\upsilon_t$ is a decreasing sequence such that $\sum_{t\ge 1} \upsilon_t=\infty$ and $\{ t_i \}_{i\ge 1}$ is an infinite set of integers with positive natural density. 

Moreover, since $\kappa_t(\mathsf{d} x|x^{\prime }) = \sfk_t(x|x^{\prime })\lambda(\mathsf{d} x)$, Definition \ref{defReshaped}(ii) yields 
\begin{equation}
\tilde \sfk_t^{\sfc,n}(x|x') = \pi_{t-1}(C_t^n) \sfk_t(x|x') + \sfr_t^n(x),  \label{eqSph0.1}
\end{equation}
where $\sfr_t^n = \frac{\sd \rho_t^n}{\sd \lambda}$ and $\tilde \sfk_t^{\sfc,n}$ are the pdf's associated to the measure $\rho_t^n$ and the reshaped kernel $\tilde \kappa_t^{\sfc,n}(\sd x| x')$, respectively. Taking \eqref{eqSph0.0-2} and \eqref{eqSph0.1} together yields 
\begin{equation}
\inf_{(x,x') \in C_{t_i}^n \times C_{t_i-1}^n} \tilde \sfk_t^{\sfc,n}(x|x') > \upsilon_{t_i} \pi_{t_i-1}(C_{t_i-1}^n) > \epsilon \upsilon_{t_i}  \label{eqSph0.2}
\end{equation}
for every $i > i_n$, with the last inequality following from the assumption \eqref{eqBigenough}.

If we recall that $\sum_{i \ge i_n} \upsilon_{t_i} = \infty$ and the fact that Assumption \ref{asUpper} implies  
\begin{equation*}
\sup_{t \ge 1; x,x'\in\mX} \tilde \sfk^{\sfc,n}(x|x') < C_0 < \infty,
\end{equation*}
then it is straightforward to combine the inequality \eqref{eqSph0.2} with expression \eqref{eqBigenough} in order to to apply Theorem \ref{thStability} and show that the PU operators $\tilde \Phi_t^{\sfc,n}$ are stable.
\end{proof}

If we put together Theorem \ref{thStableSpheres-2} and Theorem \ref{thEpsilonError}, it turns out the if we guarantee $\sup_{t\ge 1} \pi_t(\bar C_t)<\epsilon$, then the operator $\tilde \Phi_{t|0}^{\sfc,n}$ resulting from $\tilde \mS^{\sfc,n} = \{\pi_0,\tilde \kappa^{\sfc,n},g^{\sfc,n} \}$ is stable and attains uniform approximation errors over time. The lemma below provides a sufficient condition of the potentials $g$ and the transition pdf's $\sfk = \{\sfk_t\}_{t\ge 1}$ that entails $\sup_{t\ge 1} \pi_t(\bar C_t)<\epsilon$.

\begin{lemma} \label{lmSufficient}
Assume that, for every $t \ge 1$, the potential $g_t$ is positive and bounded and there are uniformly bounded conditional pdf's $\sfk_t(\cdot|\cdot)$ such that $\kappa_t(\sd x|x') = \sfk_t(x|x')\sd x$ for any $x,x'\in\mX$. If 
\begin{equation}
\lim_{n\rw \infty} \sup_{t\ge 1} \frac{
        \int_{\bar C_t^n} g\left( x_t \right) \sd x_t}
{
        \inf_{ (x_t,x_{t-1}) \in  C_t^n \times C_{t-1}^n } \sfk_t(x_t|x_{t-1})
} = 0  \label{wqw}
\end{equation}%
then we have
$\lim_{n\rw \infty } 
        \sup_{t \ge 1} \pi_t ( \bar C_t^n  ) = 0.$
\end{lemma}

\begin{proof} 
The posterior probability of the set $\bar C_t^n$ can be explicitly written in terms of $g_t$ and $\sfk_t$ as 
$$
\pi _t(\bar C_t^n) = \frac{
        \int_{\bar C_t^n} \int g_t(x_t) \sfk_t(x_{t},x_{t-1}) \pi_{t-1}(\sd x_{t-1}) \sd x_t
}{
        \int \int g_t(x_{t}) \sfk_t( x_{t},x_{t-1}) \pi_{t-1}( \sd x_{t-1} ) \sd x_t 
}
$$
and hence we can compute the upper bound
\begin{equation}
\pi _t(\bar C_t^n) \le \frac{
        \| \sfk \|_\infty \int_{\bar C_t^n} g_t( x_{t} ) \sd x_{t}
}{
         \inf_{ (x_t,x_{t-1}) \in  C_t^n \times C_{t-1}^n } \sfk_{t}(x_{t}|x_{t-1}) \int_{C_t^n} g_t(x_{t}) \sd x_t \left( 1 - \pi_{t-1}(\bar C_{t-1}^n ) \right) 
}.
\label{eqUpBo}
\end{equation}

From \eqref{wqw} we deduce that
\begin{equation*}
\lim_{n\rw \infty} \sup_{t\ge 1} \int_{\bar C_t^n} g_t(x_{t}) \sd x_{t} = 0,
\end{equation*}%
which, in turn, ensures the existence of $N>0$ such that for any $n\ge N$ we have $\int_{C_t^n} g_t( x_{t} ) \sd x_{t} \ge \frac{1}{2}$ for any $t\geq 0$. Therefore, using  \eqref{wqw} again, for any $\varepsilon >0$ there exists $N_{\varepsilon}$ such that, for every $n \ge N \vee N_\varepsilon$ and any $t\geq 1$, 
\begin{eqnarray}
\frac{
        \| \sfk \|_\infty \int_{\bar C_t^{n}} g_t( x_{t} ) \sd x_{t}
}{
         \inf_{ (x_t,x_{t-1}) \in  C_t^n \times C_{t-1}^n }\int_{C_t^n} g_t(x_{t}) \sd x_t
} &\le& 2\| \sfk \|_\infty \frac{
         \int_{\bar C_t^{n}} g_t( x_{t} ) \sd x_{t}
}{
         \inf_{ (x_t,x_{t-1}) \in  C_t^n \times C_{t-1}^n }\sfk_{t}(x_{t}|x_{t-1})
} \nonumber \\
&\le& \frac{
        \varepsilon 
}{
        ( 1+ \varepsilon )^2
}. \label{eqUpBo2}
\end{eqnarray}
Then, by an induction argument, if $\pi_0( \bar C_0^n) \le \frac{\varepsilon}{1+\varepsilon}$ then $\pi_t(\bar C_t^n) \le \frac{\varepsilon}{1+\varepsilon}$. To see this, simply combine \eqref{eqUpBo} and \eqref{eqUpBo2} to obtain
\[
\pi _t(\bar C_t^n) 
\le \frac{
        \varepsilon 
}{
        ( 1+ \varepsilon )^2
}  \times \frac{1}{1 - \pi_{t-1}(\bar C_{t-1}^n)} \nonumber \\
\le \frac{
        \varepsilon 
}{
        ( 1+ \varepsilon )^2
}  \times \frac{
        1
}{
        1 - \frac{\varepsilon}{1+\varepsilon}
} \le  \frac{\varepsilon}{1+\varepsilon}, 
\]
which completes the proof.
\end{proof}

The following statement brings together the results of this Section.

\begin{theorem}
Let $\mS$ be a model in $\mfR$, let $\sfc^n=\{ C_t^n \}$ be constructed as in Section \ref{ssCompactBalls} and let $\tilde \mS^{\sfc,n}$ be the sequence of approximate models, with operators $\tilde \Phi_{t|0}^{\sfc,n}$. If \eqref{wqw} holds, then for any constant $\epsilon>0$ (independent of $t$) there is some $n_\epsilon<\infty$ such that, for every $n>n_\epsilon$, 
\begin{itemize}
\item[(i)] $\sup_{t\ge 1} D_{tv}(\tilde \Phi_{t|0}^{\sfc,n}(\pi_0), \Phi_{t|0}(\pi_0) ) \le \epsilon$ and 
\item[(ii)] the operator $\tilde \Phi_{t|0}^{\sfc,n}$ is stable.
\end{itemize}
\end{theorem}

\begin{proof} 
Lemma \ref{lmSufficient} ensures that $\inf_{t\ge 1} \pi_t(C_t^n) > 1-\epsilon$ for sufficiently large $n$. Then, Theorem \ref{thStableSpheres-2} entails stability and Theorem \ref{thEpsilonError} guarantees that the approximation errors are uniformly bounded over time.
\end{proof}

\section{Conclusions} \label{sConclusions}

We have investigated a general scheme for the stable approximation of optimal filters generated by state space Markov models. The approximate filters are obtained by truncating the original potential functions. The construction enables us to investigate topological properties of families of optimal filters. In particular, we introduce a natural topology on the set of state space models within which the class of stable state space models form a dense set. We also show, by way of an example, that the class of models contains unstable filters and illustrate in detail how the proposed technique can be put to work in order to obtain stable approximations of these unstable filters.
This is, to the best of our knowledge, the first result regarding the topology of optimal filters.
In the last part of the paper we investigate further  approximate filters that involve not only the truncation of the original potentials, but also the modification of the corresponding Markov kernels. For such filters we investigate their stability as well as their uniform convergence over time to the original filter.


%
\appendix

%
\section{Proof of Proposition \ref{propDecompose}} \label{apDecompose}

We apply a straightforward induction argument. At time $t=0$ it is straightforward to verify that 
$$
\pi_0 = \ell_{0,\pi_0}^+ \pi_0^+ + \ell_{0,\pi_0}^- \pi_0^-,
$$
by simple inspection of Eqs. \eqref{eq4app-0} and \eqref{eqEll0}.
Let us now assume that, at time $t-1\ge 0$, the optimal filter can be decomposed as
\begin{equation}
\pi_{t-1} = \ell_{t-1,\pi_0}^+ \pi_{t-1}^+ + \ell_{t-1,\pi_0}^- \pi_{t-1}^-,
\label{eqDcp-0}
\end{equation}
where $\pi_{t-1}^+ = \Phi_{t-1|0}(\pi_0^+)$ and $\pi_{t-1}^- = \Phi_{t-1|0}(\pi_0^-)$. Then, 
\begin{eqnarray}
(f,\pi_t) &=& \frac{
        (fg_t,\kappa_t\pi_{t-1})
}{
        (g_t,\kappa_t\pi_{t-1})
} \nonumber\\
&=& \ell_{t-1,\pi_0}^+ \frac{
        (fg_t,\kappa_t\pi_{t-1}^+)
}{
        (g_t,\kappa_t\pi_{t-1})
} +  \ell_{t-1,\pi_0}^- \frac{
        (fg_t,\kappa_t\pi_{t-1}^-)
}{
        (g_t,\kappa_t\pi_{t-1})
} \label{eqDcp-1} \\
&=& \ell_{t-1,\pi_0}^+ \frac{
        (g_t,\kappa_t\pi_{t-1}^+)
}{
        (g_t,\kappa_t\pi_{t-1})
}  \frac{
        (fg_t,\kappa_t\pi_{t-1}^+)
}{
        (g_t,\kappa_t\pi_{t-1}^+)
} + \ell_{t-1,\pi_0}^- \frac{
        (g_t,\kappa_t\pi_{t-1}^-)
}{
        (g_t,\kappa_t\pi_{t-1})
}  \frac{
        (fg_t,\kappa_t\pi_{t-1}^-)
}{
        (g_t,\kappa_t\pi_{t-1}^-)
}, \label{eqDecompo-0}
\end{eqnarray}
where \eqref{eqDcp-1} is a consequence of \eqref{eqDcp-0}.  Eqs. \eqref{eqDecompo_t}, \eqref{eqRecurs1} and \eqref{eqRecurs2} follow readily from \eqref{eqDecompo-0} if we simply note that 
$$
\pi_{t-1}^+ = \Phi_{t-1|0}(\pi_0^+), \quad 
\pi_{t-1}^- = \Phi_{t-1|0}(\pi_0^-), \quad 
\pi_{t-1} = \Phi_{t-1|0}(\pi_0)
$$  
and 
$
\Phi_t(\alpha) = \frac{
        (fg_t,\kappa_t\alpha)
}{
        (g_t,\kappa_t\alpha)
}. 
$
$\QED$

%
\section{Proof of Proposition \ref{propUnstable}} \label{apUnstable}

Choose $\pi_0$ with no atoms at the origin and with support only on the positive half-line $[0,\infty)$, then choose $\bar\pi_0$ with no atoms at the origin and support only on the negative half-line $(-\infty, 0)$. Since the sequence of states $X_t$ never changes sign, it is apparent that $\pi_t^+ = \Phi_{t|0}(\pi_0^+)=\Phi_{t|0}(\pi_0)=\pi_t$ and $\pi_t^- = \Phi_{t|0}(\bar\pi_0^-)=\Phi_{t|0}(\bar\pi_0)=\bar\pi_t$, hence $\pi_t$ has support only on the positive half-line $[0,\infty)$ and $\bar\pi_t$ has support only on the negative half-line $(-\infty, 0)$ (and no atoms at the origin). Hence, the two measures are singular, i.e., $\lim_{t\rw\infty}D_{tv}(\bar\pi_t, \pi_t)=1$ and, therefore, $\Prob(A_u)=1$.

The operator $\Phi_t$ is still unstable even if we adopt the definition of weak stability (see Section \ref{ssStab}). To see that also in this case $\Prob(A_u)>0$, choose $\pi_0$ absolutely continuous with respect to the Lebesgue measure (with full support) such that $\pi_0 ((0,\infty)>\frac{1}{2}>\pi_0 ((-\infty,0))$. For example, choose $\pi_0=\mN(1,1)$, i.e., a normal distribution with mean 1 and variance 1.  

Next, choose $\bar \pi_0$ to be the measure symmetric to $\pi_0$ with respect to the origin, i.e., $\bar \pi_0(S)=\pi_0 (-S)$ for any Borel set $S$, where $-S:=\{x \in \Real: -x\in S\}$. For example, if $\pi_0=\mN(1,1)$ then $\bar \pi_0 = \mN(-1,1)$. Since both the state equation and the observation equation are symmetric w.r.t. the origin, it follows that $\bar \pi_t (S) = \pi_t(-S)$ for any Borel set $S$. Let us denote by $p_t$ and $\bar p_t$ the pdf of $\pi_t$ and $\bar \pi_t$, respectively, w.r.t. the Lebesgue measure. Then $\bar p_t(x)=p_t(-x)$ for all $x\in \mathbb R$. 

If $\Phi_t$ is weakly stable then we have $\lim_{t\rw\infty} D_{tv}(\bar\pi_t, \pi_t)=0$. This implies, in particular, that      
$$
0 = \lim_{t\rw\infty} \bar\pi_t\left((0,\infty)\right) - \pi_t\left((0,\infty)\right)
= \lim_{t\rw\infty} \pi_t\left( (-\infty, 0) \right) - \pi_t\left( (0,\infty) \right),
$$
hence $1-2\lim_{t\rw\infty}\pi_t\left((0,\infty)\right) = 0$ and, as a consequence, $\lim_{t\rw\infty}\pi_t\left((0,\infty)\right)=\frac{1}{2}$. 

\cblue{Instead of choosing an arbitrary but fixed sequence of observations $\{ Y_t=y_t \}_{t\ge 1}$, let us now assume that the observations are r.v.'s and $\Prob (A_u)=0$}, i.e., $\Phi_t^{Y_t}$ is weakly stable a.s. Then, $\lim_{t\rw\infty}D_{tv}(\bar\pi_t^{Y_{1:t}}, \pi_t^{Y_{1:t}})=0$ a.s. and, therefore, $\lim_{t\rw\infty}\pi_t^{Y_{1:t}}((0,\infty))=\frac{1}{2}$ a.s. By the bounded convergence theorem, we deduce that 
$$
\frac{1}{2} =
\lim_{t\rw\infty} \mbE\left[ \pi_t^{Y_{1:t}}\left( (0,\infty) \right) \right] =
\lim_{t\rw\infty} \mbE\left[ \Prob\left( X_t \in (0,\infty) | \mG_t \right) \right] =
\lim_{t\rw\infty}\Prob\left( X_t \in (0,\infty) \right),
$$
where $\mG_t$ is the $\sigma$-algebra generated by $Y_{1:t}$. However, the Markov kernel of the process $X_t$ does not transfer mass from the positive half-line to the negative half-line and viceversa. More precisely, if $X_0(\omega) > 0$ then $X_t(\omega) > 0$ and, similarly, $X_0(\omega) < 0$ implies $X_t(\omega) < 0$. If follows that $\Prob(X_t\ge 0)= \Prob(X_0\ge 0 )$ and, if $\Prob(X_0>0) > \frac {1}{2}$ (e.g., for $\pi_0=\mN(1,1)$) then $\Prob(X_t>0)>\frac{1}{2}$ for every $t > 0$ (and it cannot converge to $\frac{1}{2}$). This is a contradiction, hence necessarily $\Prob(A_u)>0$. $\QED$

\section{\cblue{Proof of Lemma \ref{lmReshaped}}} \label{apReshaped}

We proceed with an induction argument. At time $t=1$ we have $\tilde \xi_1^\sfc :=\tilde \kappa^\sfc_1 \pi_0 = \kappa_1\pi_0=\xi_1$ and Eq. \eqref{eqXi-Ct} holds. Moreover, for any integrable $f : \mX\mapsto\Real$ we have \begin{eqnarray}
(\Ind_{C_1}f,\pi_1) &=& 
\frac{
        (\Ind_{C_1} g_1,\xi_1)
}{
        (g_1,\xi_1)
} \times \frac{
        (\Ind_{C_1} f g_1,\xi_1)
}{
        (\Ind_{C_1} g_1, \xi_1)
} \nonumber \\
&=& \pi_1(C_1) \times \frac{
        (f g_1^\sfc,\tilde \xi_1^\sfc)
}{
        (g_1^\sfc, \tilde \xi_1^\sfc)
} =  \pi_1(C_1) (f, \tilde \pi_1^\sfc), \label{eqS3}
\end{eqnarray} 
where the first identity in \eqref{eqS3} follows from the definition $g_t^\sfc=g_t\Ind_{C_t}$ and the fact that $\tilde \xi_1^\sfc = \xi_1$. The last equality is straightforward and completes the proof for $t=1$. 

For the induction step, let us assume that 
\begin{equation}
(\Ind_{C_{t-1}} f, \pi_{t-1}) = (f, \tilde \pi_{t-1}^\sfc) \pi_{t-1}(C_{t-1})
\label{eqInductionHypo}
\end{equation}
for any integrable $f:\mX\mapsto\Real$. We evaluate the difference $(\Ind_{C_t} f, \xi_t) - (\Ind_{C_t} f, \tilde \xi_t^\sfc)$ first. We recall that $\xi_t = \kappa_t\pi_{t-1}$ and $\tilde \xi_t^\sfc = \tilde \kappa^\sfc_t \tilde \pi_{t-1}^\sfc$, hence,
\begin{eqnarray}
(\Ind_{C_t} f, \xi_t) - (\Ind_{C_t} f, \tilde \xi_t^\sfc) &=& \left( (\Ind_{C_t} f, \kappa_t), \pi_{t-1}\right) - \left( (\Ind_{C_t} f, \tilde \kappa^\sfc_t), \tilde \pi_{t-1}^\sfc\right) \nonumber\\
&=& \left(
        (\Ind_{C_t} f, \kappa_t)\Ind_{C_{t-1}},\pi_{t-1}
\right) +  \left(
        (\Ind_{C_t} f, \kappa_t) \Ind_{\bar C_{t-1}},\pi_{t-1}
\right) \nonumber \\
&& -  \pi_{t-1}(C_{t-1}) \left(
        (\Ind_{C_t} f, \kappa_t), \tilde \pi_{t-1}^\sfc
\right) - \left( (\Ind_{C_t} f, \rho_t), \tilde \pi_{t-1}^\sfc \right),
\label{eqDecomp1}
\end{eqnarray}
where the last equality is obtained by substituting $\tilde \kappa_t^\sfc = \pi_{t-1}(C_{t-1})\kappa_t + \rho_t$. However, $\left( (\Ind_{C_t} f, \rho_t), \tilde \pi_{t-1}^\sfc \right) =  (\Ind_{C_t} f, \rho_t)$, hence \eqref{eqDecomp1} becomes
\begin{eqnarray}
(\Ind_{C_t} f, \xi_t) - (\Ind_{C_t} f, \tilde \xi_t^\sfc) &=& \left(
        (\Ind_{C_t} f, \kappa_t)\Ind_{C_{t-1}},\pi_{t-1}
\right) +  \left(
        (\Ind_{C_t} f, \kappa_t) \Ind_{\bar C_{t-1}},\pi_{t-1}
\right) \nonumber \\
&& -  \pi_{t-1}(C_{t-1}) \left(
        (\Ind_{C_t} f, \kappa_t), \tilde\pi_{t-1}^\sfc
\right) - (\Ind_{C_t} f, \rho_t).
\label{eqDecomp1-4}
\end{eqnarray}

Let us now compare the first and third terms in Eq. \eqref{eqDecomp1-4}. If we define the function
$\sff_t(x) := \int \Ind_{C_t}(x') f(x') \kappa_t(\sd x' | x)$
then it is straightforward to see that the first term on the r.h.s. of \eqref{eqDecomp1-4} can be rewritten as
\begin{equation}
\left(
        (\Ind_{C_t} f, \kappa_t)\Ind_{C_{t-1}},\pi_{t-1}
\right) = (\Ind_{C_{t-1}} \sff_t, \pi_{t-1}),
\label{eqT1}
\end{equation}
and we obtain the same expression for the third term, namely,
\begin{equation}
\pi_{t-1}(C_{t-1}) \left(
        (\Ind_{C_t} f, \kappa_t), \tilde\pi_{t-1}^\sfc
\right) = \pi_{t-1}(C_{t-1}) (\sff_t, \tilde \pi_{t-1}^\sfc) 
= (\Ind_{C_{t-1}} \sff_t, \pi_{t-1}),
\label{eqT3-3}
\end{equation}
where the second equality follows from the induction hypothesis \eqref{eqInductionHypo}. 

We are now left with the comparison of the second and fourth terms in \eqref{eqDecomp1}. For the second term, it is straightforward to see that
\begin{equation}
\left(
        (\Ind_{C_t} f, \kappa_t) \Ind_{\bar C_{t-1}},\pi_{t-1}
\right) = (\Ind_{\bar C_{t-1}} \sff_t, \pi_{t-1}),
\label{eqT2}
\end{equation}
while from the definition of $\rho_t(\sd x)$ in Eq. \eqref{eqDefRho} and $\sff_t$ above 
we obtain an identical expression for the fourth term, i.e.,
\begin{equation}
(\Ind_{C_t} f, \rho_t) = ( \Ind_{\bar C_{t-1}} \sff_t, \pi_{t-1}). \label{eqT4}
\end{equation} 
If we substitute \eqref{eqT1}, \eqref{eqT3-3}, \eqref{eqT2} and \eqref{eqT4} into Eq. \eqref{eqDecomp1-4} 
we arrive at the equality $(\Ind_{C_t} f, \xi_t) = (\Ind_{C_t} f, \tilde \xi_t^\sfc)$ in Eq. \eqref{eqXi-Ct}.

To conclude the proof, we repeat the argument of time $t=1$ for the filter $\pi_t$, namely,
\begin{eqnarray}
(\Ind_{C_t}f,\pi_t) &=& \frac{
        (\Ind_{C_t}fg_t,\xi_t)
}{
        (g_t,\xi_t)
} = \frac{
        (\Ind_{C_t} g_t, \xi_t)
}{
        (g_t, \xi_t)
} \times \frac{
        (\Ind_{C_t} f g_t, \xi_t)
}{
        (\Ind_{C_t} g_t, \xi_t)
} \nonumber \\ 
&=& \pi_t(C_t) \times \frac{
        (f g_t^\sfc, \tilde\xi_t^\sfc)
}{
        (g_t^\sfc, \tilde \xi_t^\sfc)
} = \pi_t(C_t) (f, \tilde \pi_t^\sfc) \label{eqS3-2}
\end{eqnarray} 
where \eqref{eqS3-2} follows from \eqref{eqXi-Ct} and the fact that $g_t^\sfc = \Ind_{C_t}g_t$. 
$\QED$

\bibliographystyle{siamplain}
\bibliography{bibliografia}

\end{document}